\documentclass[traditabstract,longauth]{aa} 
\usepackage{graphicx}
\usepackage{color}
\usepackage{txfonts}
\begin{document}

   \title{The Gaia-ESO Survey: calibration strategy\thanks{Based on data
products from observations made with ESO Telescopes at the La Silla Paranal
Observatory under programme IDs 188.B-3002 and 193.B-0936.}}

   \author{E.~Pancino\inst{1,2}
          \and
          C.~Lardo\inst{3}
          \and
          G.~Altavilla\inst{4}
          \and
          S.~Marinoni\inst{5,2}
          \and
          S.~Ragaini\inst{4}
          \and
          G.~Cocozza\inst{4}
          \and
          M.~Bellazzini\inst{4}
          \and
          E.~Sabbi\inst{6}
          \and
          M.~Zoccali\inst{7,8}
          \and
          P.~Donati\inst{4,9}
          \and
          U.~Heiter\inst{10}
          \and
          S.~E.~Koposov\inst{11}
          \and
          R.~Blomme\inst{12}
          \and
          T.~Morel\inst{13}
          \and
          S.~S\'imon-D\'{i}az\inst{14,15}
          \and
          A.~Lobel\inst{12}
          \and
          C.~Soubiran\inst{16}
          \and
          J.~Montalban\inst{17,27}
          \and
          M.~Valentini\inst{18}
          \and
          A.~R.~Casey\inst{11}
          \and
          S.~Blanco-Cuaresma\inst{19}
          \and
          P.~Jofr\'e\inst{11,20}
          \and
          C.~C.~Worley\inst{11}
          \and
          L.~Magrini\inst{1}
          \and
          A.~Hourihane\inst{11}
          \and
          P.~Fran\c cois\inst{21,39}
          \and
          S.~Feltzing\inst{22}
          \and
          G.~Gilmore\inst{11}
          \and
          S.~Randich\inst{1}
          \and
          M.~Asplund\inst{23}
          \and
          P.~Bonifacio\inst{21}
          \and
          J.~E.~Drew\inst{24}
          \and
          R.~D.~Jeffries\inst{25}
          \and
          G.~Micela\inst{26}
          \and
          A.~Vallenari\inst{27}
          \and
          E.~J.~Alfaro\inst{28}
          \and
          C.~Allende Prieto\inst{14,15}
          \and
          C.~Babusiaux\inst{21}
          \and
          T.~Bensby\inst{22}
          \and
          A.~Bragaglia\inst{4}
          \and
          E.~Flaccomio\inst{26}
          \and
          N.~Hambly\inst{29}
          \and
          A.~J.~Korn\inst{10}
          \and
          A.~C.~Lanzafame\inst{30,34}
          \and
          R.~Smiljanic\inst{31}
          \and
          S.~Van Eck\inst{32}
          \and
          N.~A.~Walton\inst{11}
          \and
          A.~Bayo\inst{33}
          \and
          G.~Carraro\inst{17}
          \and
          M.~T.~Costado\inst{28}
          \and
          F.~Damiani\inst{26}
          \and
          B.~Edvardsson\inst{10}
          \and
          E.~Franciosini\inst{1}
          \and
          A.~Frasca\inst{34}
          \and
          J.~Lewis\inst{11}
          \and
          L.~Monaco\inst{35}
          \and
          L.~Morbidelli\inst{1}
          \and
          L.~Prisinzano\inst{26}
          \and
          G.~G.~Sacco\inst{1}
          \and
          L.~Sbordone\inst{8,7,34}
          \and
          S.~G.~Sousa\inst{37}
          \and
          S.~Zaggia\inst{27}
          \and
          A.~Koch\inst{38}
          }

   \institute{INAF -- Osservatorio Astrofisico di Arcetri, Largo E. Fermi 5, 
              50125 Firenze, Italy, \email{pancino@arcetri.inaf.it}
         \and
              ASI Science Data Center, Via del Politecnico s/n, 00133 Roma
         \and
              Astrophysics Research Institute, Liverpool John Moores University,
              146 Brownlow Hill, Liverpool L3 5RF, United Kingdom
         \and 
              INAF -- Osservatorio Astronomico di Bologna, Via Ranzani 1, 
              40127 Bologna, Italy
         \and 
              INAF -- Osservatorio Astronomico di Roma, Via Frascati 33, 
              00040, Monte Porzio Catone, Roma, Italy 
         \and             
              Space Telescope Science Institute, 3700 San Martin Drive,
              Baltimore, MD 21218, US            
         \and
              Instituto de  Astrof\'isica, Pontificia Universidad  Cat\'olica de
              Chile, Av. Vicu\~na Mackenna 4860, 782-0436 Macul, Santiago, Chile
         \and
              Millennium  Institute of Astrophysics, Av. Vicu\~na Mackenna 4860,
              782-0436 Macul, Santiago, Chile
         \and
              Dipartimento di Fisica e Astronomia, Alma Mater Studiorum,
              Universit\`a di Bologna, Viale Berti Pichat 6/2, 40127, Bologna,
              Italy
         \and
              Department of Physics and Astronomy, Uppsala University, Box 516,
              SE-751 20 Uppsala, Sweden
         \and
              Institute of Astronomy, University of Cambridge, Madingley Road, 
              Cambridge CB3 0HA, United Kingdom
         \and
              Royal Observatory of Belgium, Ringlaan 3, 1180, Brussels, Belgium
         \and
              Institut d'Astrophysique et de G\'eophysique, Universit\'e de Li\`ege, 
              All\'ee du 6 Ao\^ut, B\^at. B5c, 4000 Li\`ege, Belgium
         \and
              Instituto de Astrof\'{i}sica de Canarias, E-38200 La Laguna, 
              Tenerife, Spain
         \and
              Departamento de Astrof\'{i}sica, Universidad de La Laguna,
              E-38205 La Laguna, Tenerife, Spain
         \and
              Laboratoire d'astrophysique de Bordeaux, Univ. Bordeaux, CNRS,
              B18N, all\'e Geoffroy Saint-Hilaire, 33615 Pessac, France
         \and
              Dipartimento di Fisica e Astronomia, Universit\`a di Padova,
              Vicolo dell'Osservatorio 2, 35122 Padova, Italy
         \and
              Leibnitz Institute f\"ur Astrophysics (AIP), An der Sternwarte 16,
              14482 Potsdam, Germany
         \and
              Observatoire de Gen\`eve, Universit\'e de Gen\`eve, CH-1290 Versoix,
              Switzerland
         \and
              N\'ucleo de Astronom\'ia, Facultad de Ingenier\'ia, Universidad 
              Diego Portales, Av. Ejercito 441, Santiago, Chile
         \and
              GEPI, Observatoire de Paris, PSL Research University, CNRS, Univ.
              Paris Diderot, Sorbonne Paris Cit{\'e}, 5 Place Jules Janssen,
              92190 Meudon, France 
         \and
              Lund Observatory, Department of Astronomy and Theoretical Physics,
              Box 43, SE-221 00 Lund, Sweden
         \and
              Research School of Astronomy \& Astrophysics, Australian National 
              University, Cotter Road, Weston Creek, ACT 2611, Australia     
         \and
              Centre for Astrophysics Research, STRI, University of 
              Hertfordshire, College Lane Campus, Hatfield AL10 9AB, United 
              Kingdom
         \and
              Astrophysics Group, Research Institute for the Environment, 
              Physical Sciences and Applied Mathematics, Keele University, Keele,
              Staffordshire ST5 5BG, United Kingdom
         \and
              INAF - Osservatorio Astronomico di Palermo, Piazza del Parlamento 
              1, 90134, Palermo, Italy
         \and
              INAF - Padova Observatory, Vicolo dell'Osservatorio 5, 35122 
              Padova, Italy
         \and
              Instituto de Astrof\'{i}sica de Andaluc\'{i}a-CSIC, Apdo. 3004, 
              18080, Granada, Spain
         \and
              Institute of Astronomy, University of Edinburgh, Blackford Hill, 
              Edinburgh EH9 3HJ, United Kingdom
         \and
              Dipartimento di Fisica e Astronomia, Sezione Astrofisica, 
              Universit\`{a} di Catania, via S. Sofia 78, 95123, Catania, Italy
         \and
              Nicolaus Copernicus Astronomical Center, Polish Academy of
              Sciences, ul. Bartycka 18, 00-716, Warsaw, Poland
         \and
              Institut d'Astronomie et d'Astrophysique, Universit\'{e} libre de 
              Brussels, Boulevard du Triomphe, 1050 Brussels, Belgium
         \and
              Instituto de F\'{i}sica y Astronom\'{i}a, Universidad de 
              Valpara\'{i}so, Chile
         \and
              INAF - Osservatorio Astrofisico di Catania, via S. Sofia 78, 95123,
              Catania, Italy
         \and
              Departamento de Ciencias Fisicas, Universidad Andres Bello, 
              Republica 220, Santiago, Chile
         \and
              European Southern Observatory, Alonso de Cordova 3107 Vitacura, 
              Santiago de Chile, Chile
         \and
              Instituto de Astrof\'isica e Ci\^encias do Espa\c{c}o, 
              Universidade do Porto, CAUP, Rua das Estrelas, 4150-762 Porto, 
              Portugal
         \and
              Physics Department, Lancaster University, Lancaster LA1 4YB, UK
         \and
              UPJV, Universit\'e de Picardie Jules Verne, 33 Rue St Leu, F-80080
              Amiens, France
              }          
   \date{Received September 15, 1996; accepted March 16, 1997}

 
  \abstract
  {The Gaia-ESO survey (GES) is now in its fifth and last year of observations,
  and has already produced tens of thousands of high-quality spectra of stars in
  all Milky Way components. This paper presents the strategy behind the
  selection of astrophysical calibration targets, ensuring that all GES results
  on radial velocities, atmospheric parameters, and chemical abundance ratios
  will be both internally consistent and easily comparable with other literature
  results, especially from other large spectroscopic surveys and from Gaia. The
  calibration of GES is particularly delicate because of: {\em (i)} the large
  space of parameters covered by its targets, ranging from dwarfs to giants,
  from O to M stars, and with a large range of metallicities, as well as
  including fast rotators, emission line objects, stars affected by veiling and
  so on; {\em (ii)} the variety of observing setups, with different wavelength
  ranges and resolution; and {\em (iii)} the choice of analyzing the data with
  many different state-of-the art methods, each stronger in a different region
  of the parameter space, which ensures a better understanding of systematic
  uncertainties. An overview of the GES calibration and homogenization strategy
  is also given, along with some examples of the usage and results of
  calibrators in GES iDR4 -- the fourth internal GES data release, that will
  form the basis of the next GES public data release. The agreement between GES
  iDR4 recommended values and reference values for the calibrating objects are
  very satisfactory. The average offsets and spreads are generally compatible
  with the GES measurement errors, which in iDR4 data already meet the
  requirements set by the main GES scientific goals.}


   \maketitle
%

\section{Introduction}
\label{sec-intro}

The detailed study of the Milky Way (MW) as a galaxy has emerged as a central
field in modern astrophysics and is currently attracting much attention, not the
least thanks to the launch of the Gaia ESA space mission in December 2013
\citep{gaia1,gaia2,lindegren96,mignard05,gaia3}. For in-depth studies of the
properties of the stellar populations in the MW, high-multiplex spectroscopy of
sufficient resolution is required to obtain radial velocities (RV), stellar
astrophysical parameters (AP), and elemental abundances for large numbers of
stars \citep{freeman02,bland10}. Several new instruments have been designed
around this idea \citep[including HERMES, 4MOST, and
WEAVE;][]{barden10,dejong14,balcells10} and several spectroscopic surveys are
ongoing or planned with this goal in mind \citep[for example RAVE, APOGEE, GALAH,
and LEGUE;][]{kordopatis13,majewski15,galah,legue}. All these surveys will study
millions of stars, but they will adopt different selection criteria, instrumental
setups, and data analysis methods. 

The Gaia-ESO public spectroscopic survey \citep[GES,][]{gilmore12,randich12}
started operations at the end of 2011, with the goal of exploring all components
of the MW in a complementary way to Gaia. GES uses the FLAMES optical
spectrograph \citep{flames} at the ESO (European Southern Observatory) VLT (Very
Large  Telescope), in Medusa combined mode, where 6 to 8 fibers are used by UVES
with a resolution of R=$\lambda/\Delta\lambda\simeq$47\,000, and 132 fibers are
used by GIRAFFE, with R$\simeq$16\,000--25\,000, depending on the wavelength
range chosen (see Table~\ref{tab-setup} for a list of the GES observing setups
used). GES is measuring RVs and derive APs and chemical abundances of several
elements for $\sim$10$^5$ stars, focussing on relatively faint stars
(mainly V$>$16~mag), for which Gaia will not be able to provide accurate RVs
and abundances. GES data have their own outstanding scientific and legacy value,
but together with the Gaia data they will provide extremely detailed 6D space
information (position, distance, and 3D motions), combined with astrophysical
information, for a representative sample of MW stars. 

\begin{table}
\caption{FLAMES instrumental setups used in the Gaia-ESO Survey, with the number
of individual stars analysed in iDR4 for each setup. The official ESO setup data
presented here refer to the period covered by GES iDR4 observations, i.e., before
August 2014.}            
\label{tab-setup}      
\centering          
\begin{tabular}{l l c c c r}     
\hline\hline       
Instrument & Setup              & $\rm{\lambda_{min}}$ & $\rm{\lambda_{max}}$ & R & iDR4 \\ 
           &                    &  (\AA)  & (\AA) & ($\lambda/\Delta\lambda$) \\ 
\hline   
UVES    & 520\tablefootmark{a,d}  & 4140  &  6210  & 47000 &   337 \\
UVES    & 580\tablefootmark{b,d}  & 4760  &  6840  & 47000 &  3281 \\
UVES    & 860\tablefootmark{c}    & 6600  & 10600  & 47000 &   --- \\
GIRAFFE & HR3\tablefootmark{a,d}  & 4033  &  4201  & 24800 &   822 \\
GIRAFFE & HR4\tablefootmark{a,e}  & 4188  &  4297  & 24000 &   --- \\
GIRAFFE & HR5A\tablefootmark{a,d} & 4340  &  4587  & 18470 &   823 \\
GIRAFFE & HR6\tablefootmark{a,d}  & 4538  &  4759  & 20350 &   806 \\
GIRAFFE & HR9B\tablefootmark{d}   & 5143  &  5356  & 25900 &  2243 \\
GIRAFFE & HR10\tablefootmark{f}   & 5339  &  5619  & 19800 & 29215 \\
GIRAFFE & HR14A\tablefootmark{a,d}& 6308  &  6701  & 17740 &  683 \\
GIRAFFE & HR15N\tablefootmark{d}  & 6470  &  6790  & 17000 & 19431 \\
GIRAFFE & HR21\tablefootmark{f}   & 8484  &  9001  & 16200 & 31649 \\
\hline \hline                                               
\multicolumn{6}{l}{$^{(a)}$Mostly used for OBA stars (WG13).}\\
\multicolumn{6}{l}{$^{(b)}$Mostly used for FGK stars (WG10, WG11, WG12).}\\
\multicolumn{6}{l}{$^{(c)}$Used for benchmark stars (legacy value only, no
analysis).}\\
\multicolumn{6}{l}{$^{(d)}$Used for OCs; HR09B is generally used for stars of
type A}\\
\multicolumn{6}{l}{and hotter, while HR15N is used for stars of type F and cooler.}\\
\multicolumn{6}{l}{$^{(e)}$Not in iDR4, introduced only recently.}\\
\multicolumn{6}{l}{$^{(f)}$Used for MW field stars.}\\
\end{tabular}                                         
\end{table}

Stellar spectroscopic surveys require specific calibrators, to allow for
meaningful comparisons with other literature studies and spectroscopic surveys,
but also for internal homogenization purposes. GES has chosen to invest a
significant effort on calibrations, because of the large variety of stellar
targets, and consequently of observational setups and analysis methods. Of
course, the calibration objects do not serve only to assess the internal
consistency, but also to allow for external comparisons with other large surveys
and with Gaia. This will maximize their legacy value and provide a rich reference
dataset for future inter-survey calibrations.

   \begin{figure}
   \centering
   \includegraphics[width=\columnwidth]{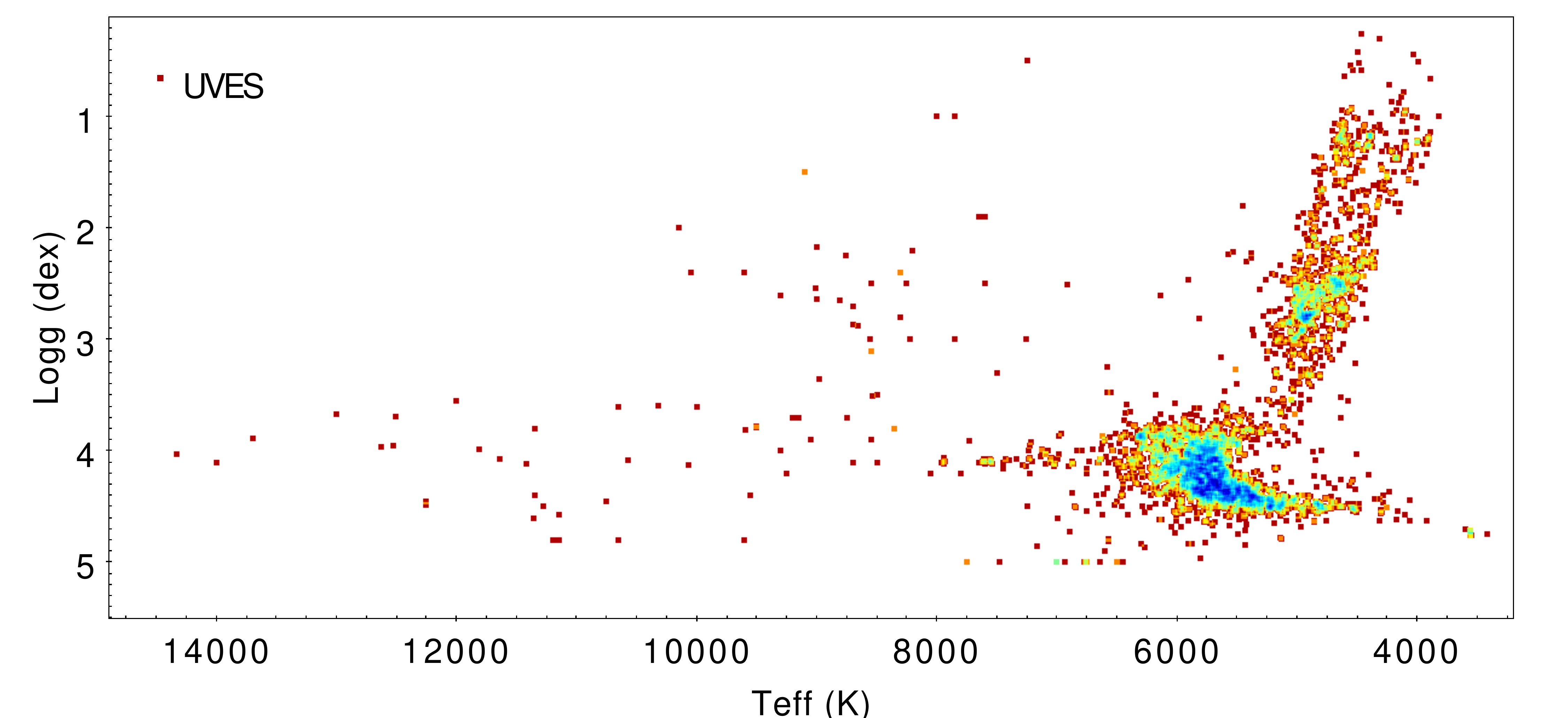}\\
   \includegraphics[width=\columnwidth]{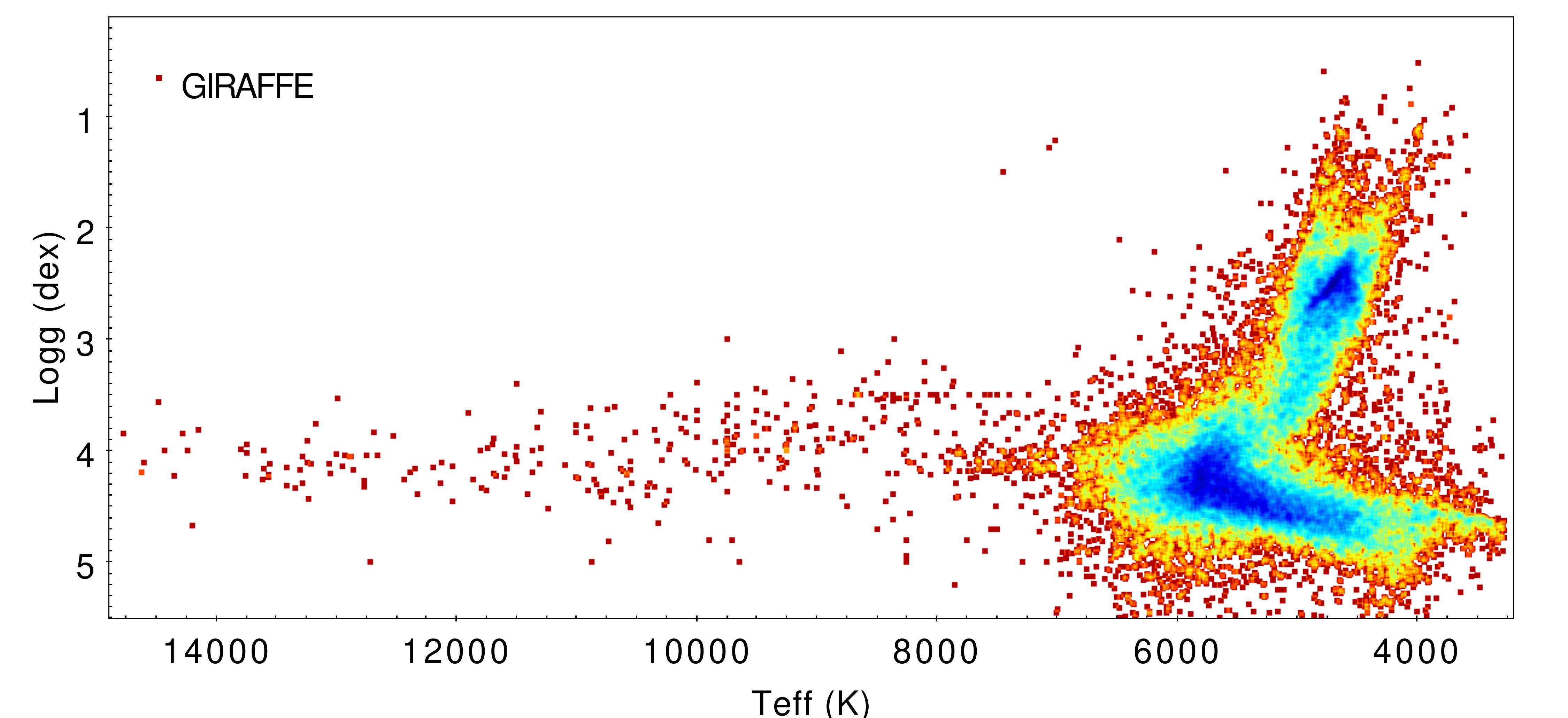}
      \caption{Parameter coverage of GES iDR4 stars. The top panel shows stars
      observed with UVES and the bottom one with GIRAFFE. The color-scale refers
      to the density of points (red is low density while blue is high). A long
      tail of hot stars extending to T$_{\rm{eff}}$>14\,000~K was cut for plot
      readability.} 
         \label{fig-params}
   \end{figure}

In this paper we describe the GES calibration needs, the calibrating targets
selection and observation processes, and the various uses and purposes of the
chosen calibrators in the framework of the GES data analysis. We use the GES iDR4
data\footnote{GES iDR4 is the fourth internal data release, where a large
fraction of the data obtained before the end of August 2014 were re-analyzed
homogeneously, taking into account the lessons learned in the previous internal
releases. GES iDR4 will also form the basis of the next GES public data release
through the ESO Phase3 portal for public surveys, which is expected in Autumn
2016.} to illustrate how the calibrators are employed in GES, and with which
results. The paper is organized as follows: in Section~\ref{sec-strat} we discuss
the general basis and implementation of the GES calibration strategy; the
following sections discuss various types of calibrators like RV standards
(Section~\ref{sec-basic}); open and globular clusters (OC and GC, respectively,
Section~\ref{sec-clusters}); benchmark stars (Section~\ref{sec-bench});
astroseismologic constraints (Section~\ref{sec-corot}). In
Section~\ref{sec-concl} we present our summary and conclusions. 


\section{GES calibration needs and strategy}
\label{sec-strat}

The broad scientific goal of GES is to survey all MW components, including the
disk(s), the bulge, the halo, with special attention to the Solar neighborhood,
that will be studied by Gaia in extreme detail \citep{gilmore12}. GES includes OCs
of all ages, excluding only those that are still embedded \citep{randich12}, to
study their internal properties and evolution, and as tracers of the thin disk
population. 

\begin{figure}
   \centering
   \includegraphics[width=\columnwidth]{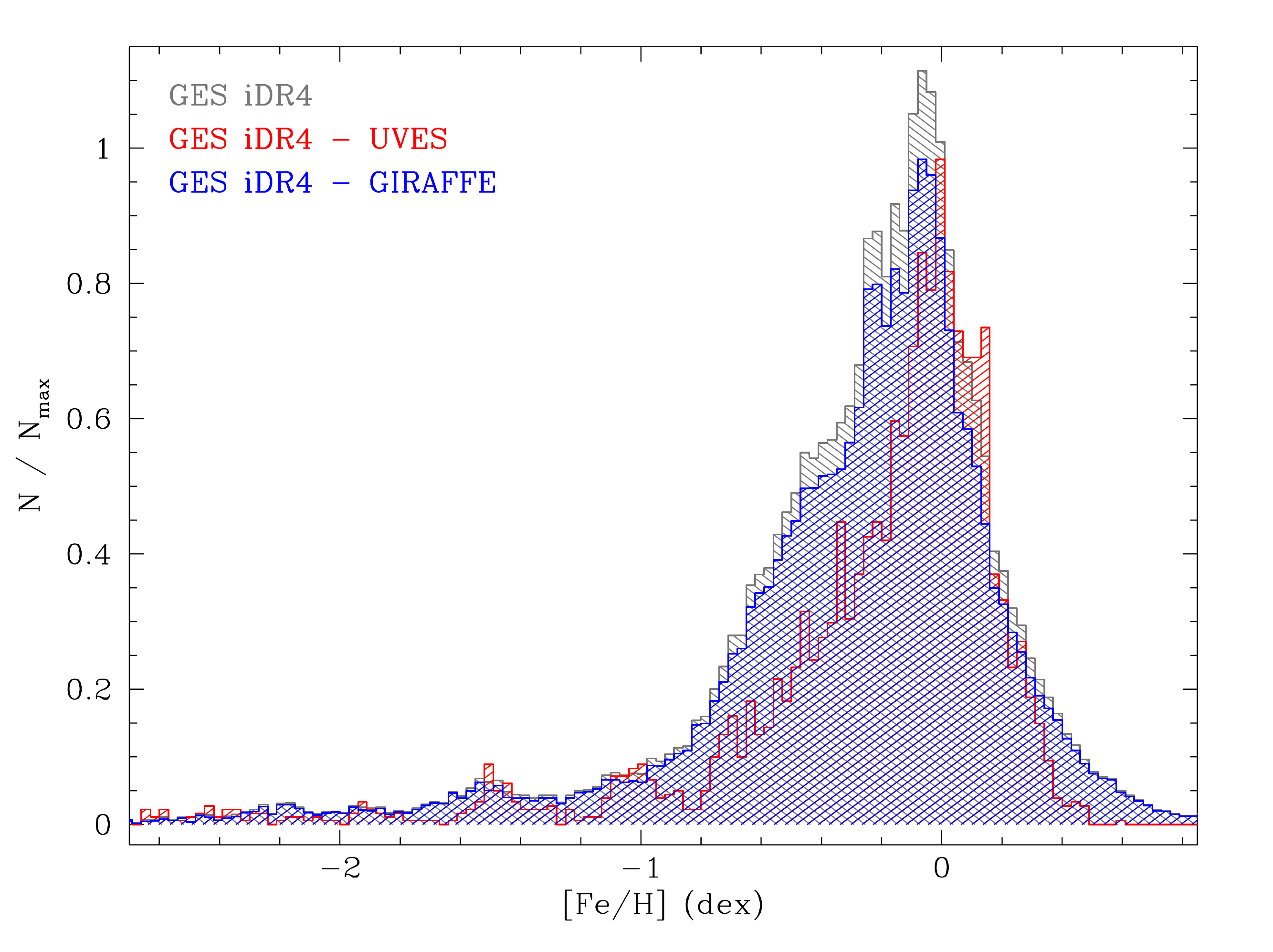}
      \caption{Metallicity distribution of GES iDR4 targets as a whole
      (grey-shaded histogram), and of the UVES (red-shaded) and GIRAFFE
      (blue-shaded) targets in iDR4. The histogram of the whole sample was
      normalized differently for clarity.} 
         \label{fig-gesmet}
   \end{figure}

As a result, GES targets cover a wide range of properties, from dwarfs to giants,
from O to M stars, and with a wide range of metallicities and abundance patterns.
Figure~\ref{fig-params} shows the parameter space coverage of the 54\,530 iDR4
GES targets for which recommended parameters\footnote{Here and in the rest of the
paper, the {\em recommended} values, APs, RVs, or abundances are the final values
produced by GES after the whole homogenization procedure.} were produced. The
corresponding [Fe/H] distribution is presented in Figure~\ref{fig-gesmet}. {\em
As a first obvious requirement, GES calibrators must cover adequately this wide
range of properties}.

The analysis of the stellar spectra obtained by GES has been organized in a set
of Working Groups (WGs). The characteristics of each WG are described in detail
elsewhere, but in short: WG10 deals with the GIRAFFE analysis of FGK stars
(Recio-Blanco et al., in preparation), WG11 with the UVES analysis of FGK stars
\citep{smiljanic14}, WG12 with the analysis of pre-MS and of cool stars
\citep{lanzafame15}, and WG13 with the analysis of hot stars (Blomme et al., in
preparation). Within each WG, almost all of the state-of-the-art methodologies --
appropriate for different objects -- are implemented and applied by various
research groups called {\em abundance analysis nodes}. They cover various
methodologies, from full spectral synthesis to classical EW (Equivalent Width)
techniques, and using a variety of abundance computation codes. Some are more
suited to deal with specific stellar properties like for example stellar rotation
or veiling. Others were designed for accurate measurements of specific features,
for example lithium or the H$_{\alpha}$ line. More details on the individual
nodes abundance analysis methods can be found in the above cited papers,
describing the WG analysis. This is a major strength of GES, because it allows
for method intercomparisons that are extremely instructive on the strengths,
weaknesses and applicability ranges of each method, and for a deep knowledge of
systematic errors. However, this complexity of the data analysis places another
strong requirement on the calibration strategy: {\em that an adequate number of
calibrating objects fall also into those regions of the parameter space
that are analyzed by more than one WG and node.}

\begin{figure}
   \centering
   \includegraphics[width=\columnwidth]{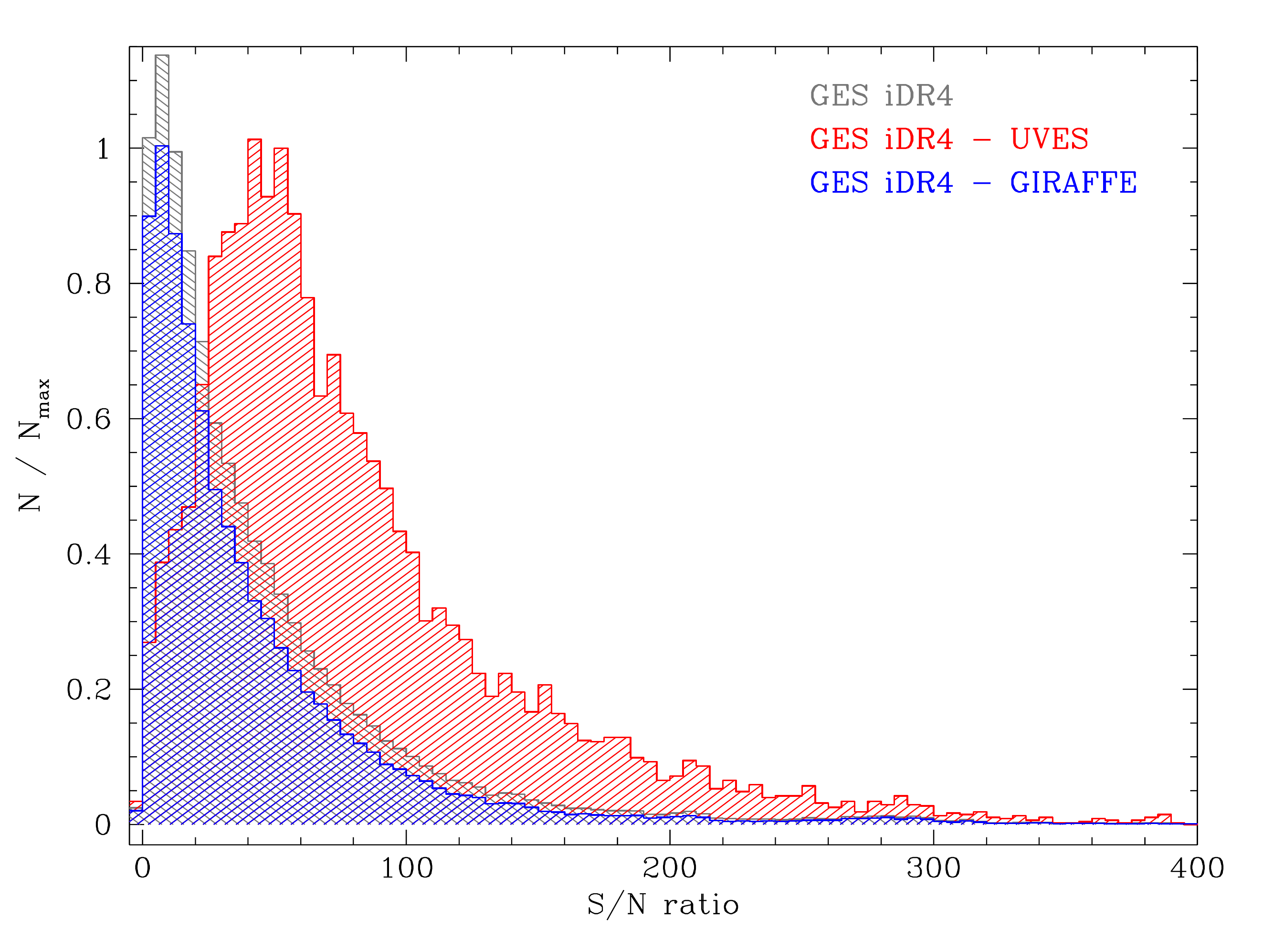}
      \caption{Histogram of the S/N ratio distribution for individual spectra in GES
      iDR4 (grey-shaded histogram), and of UVES (red-shaded) and GIRAFFE
      (blue-shaded) individual spectra. The whole iDR4 sample is normalized
      differently for clarity.} 
         \label{fig-snr}
   \end{figure}

Finally, as a natural consequence of the large variety of science targets and
methods, the observing strategy relies on several different observing setups,
that are appropriate for different types of objects and are summarized in 
Table~\ref{tab-setup}. Also, depending on the science goal (focus on RVs or on
chemical abundances), a wide range of S/N ratios were obtained, as shown in
Figure~\ref{fig-snr}. This places another requirement on GES calibrations: {\em
an adequate number of (calibrating) objects need to be observed with more than
one setup and with a range of S/N ratios}.

All the calibration requirements described above ensure that GES is both
internally consistent with respect to the different methods, objects, and
observational setups, and easily comparable with other literature results.
Therefore, a good fraction of the calibrators need to be well studied objects
with reliable reference parameters and abundances. It is also desirable that
some of the calibrators are observed by other large surveys as well, to enhance
the legacy value of GES. The ensemble of all the internal and external
calibration procedures in GES is referred to as {\em homogenisation}.

\subsection{GES analysis workflow}

\begin{figure}
   \centering
   \includegraphics[width=\hsize]{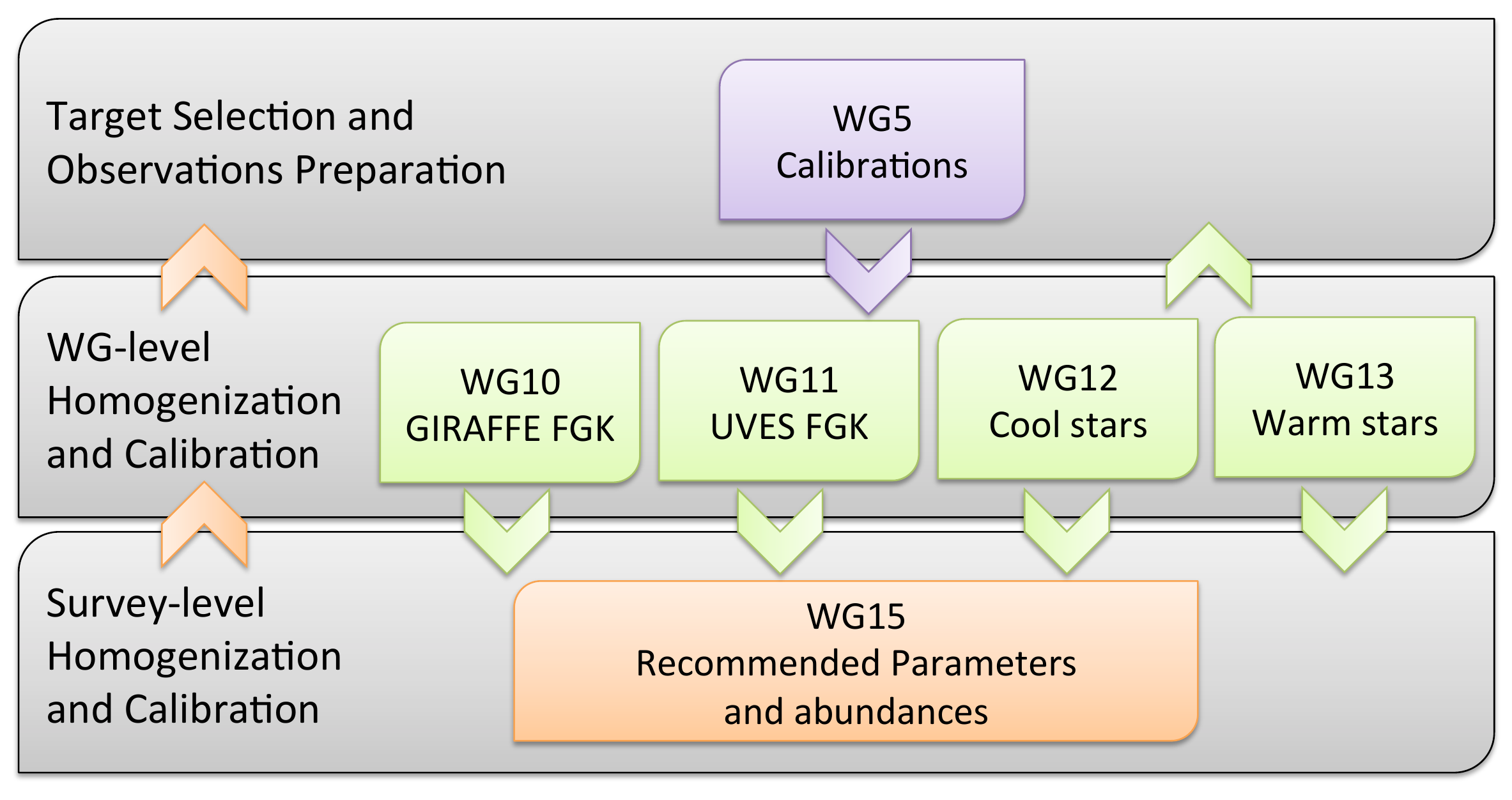}
      \caption{The iterative GES calibration and homogenisation process. Arrows
      mark the flow of information from target selection (described in this
      paper), to abundance analysis, and production of recommended parameters and
      abundances. Between and during abundance analysis cycles, feedback is
      provided by the downstream layers, to refine both the calibration
      observations and the analysis strategy. Only those WGs that make use of
      calibrators are indicated in this figure.}
        \label{fig-flow}
   \end{figure}

GES data analysis proceeds in cycles, also called internal data releases (iDR).
Within each cycle, the survey calibration and homogenisation is organized in
three logical layers, as illustrated in Figure~\ref{fig-flow}. The starting one,
coordinated by WG5, takes care of selecting the appropriate calibrating objects
and of preparing their observations, which is the main topic of the present
paper. In a second layer, appropriate calibrators are used by the WGs to compare
and combine the node-level APs and abundances into WG-level recommended
parameters. Finally, in different stages along each cycle, WG15 performs a
homogenisation of the WG-level results, to provide survey-level recommended RVs,
APs, and chemical abundance ratios. 

Internal consistency among abundace analysis nodes is facilitated as much as 
possible\footnote{This was not possible in all cases; for example, the hot stars
abundance nodes obviously relied on a different set of atmospheric models.} by
the use of a common set of atmospheric models \citep[MARCS, see,][]{marcs}, a
common linelist \citep{heiter15a}, and a common grid of synthetic spectra, based
upon the one by \citet{delaverny12}. For the first processing cycles, up to iDR3,
the homogenisation was carried out in a limited, exploratory way, based mostly on
benchmark stars. During iDR4, the first full homogenisation took place at all
levels, making use of all the observed calibrators and of new homogenisation
algorithms. This effort provided important feedback on the calibration strategy,
finalizing the calibrators selection strategy and the planning of the remaining
calibration observations. The detailed homogenization procedure and algorithms
are described in a companion paper (Hourihane et al., in preparation, hereafter
H17). 

\subsection{GES calibrator types and observing strategy}

The GES calibrators fall into a few main groups, described more in detail in
dedicated sections, summarized here below. The list of iDR4 calibrators
used in this paper is in Table~\ref{tab:cnames}.

\begin{itemize}
\item{{\em Basic calibrations} in GES are mostly related to RV standard star
observations, as described in Section~\ref{sec-basic}.}
\item{In GES, we extend the set of calibrating objects by including also {\em
benchmark stars}, and in particular the Gaia FGK benchmark stars
\citep{heiter15b}. They are carefully selected, well-studied stars, for which
T$_{\rm{eff}}$ and log$g$ were derived as independently from spectroscopy as
possible (i.e., based on interferometric diameters, parallaxes, and so on, see
Section~\ref{sec-bench} for more details). As such, they are good {\em absolute
calibrators} of the parameters (i.e., for the accuracy), and useful references
for the abundances.}
\item{Like many other spectroscopic surveys, GES observes many stars belonging
to OCs and GCs, as described more in detail in Section~\ref{sec-clusters}. These
calibrators are quite powerful for checking the internal consistency of the
abundance analysis (i.e., for the precision), as well as providing a relatively
robust external reference for the abundance scale and AP determination.}
\item{Collaborations are also ongoing with the CoRoT and Kepler teams, to obtain
accurate log$g$ reference values for large samples of giant stars, as described
in Section~\ref{sec-corot}.}
\end{itemize}

The general idea behind the observing strategy is the following. For {\em
internal} calibrations, each object should be observed with all the setups used
by the different groups that will attempt a meaningful analysis of that object.
For example, calibrators that can be in principle analysed by both OBA and FGK
star experts should be observed with the setups adopted in GES for OBA stars
(HR3, HR4\footnote{This setup was introduced later to improve the log$g$
determination for hot stars, so it was not employed for iDR4.}, HR5, HR6, and
HR14) and FGK stars (HR9B, HR10, HR15N, HR21). In another example, OCs contain
stars with properties overlapping those of the MW field part of the survey. To
ensure that both OCs and field stars are analyzed consistently, a set of
calibrating OCs should be observed with both the cluster (HR9B and HR15N) and
the field (HR10 and HR21) GIRAFFE setups. More details on the typically adopted
setups for each calibration type can be found in the following sections and in
Table~\ref{tab-setup}.

\begin{table}
\caption{List of GES iDR4 calibrators used in this paper. The list is
published in its entirety in the electronic version of the paper, and at CDS. It
can be used to select the iDR4 calibrators from the upcoming ESO Phase 3 public
release. Here we show a portion to illustrate its contents. The columns contain:
{\em (1)} the GES unique identifier of each star (the CNAME), based on the object
sexagesimal coordinates; {\em (2)} the calibration type, that can be GC or OC for
clusters, RV for radial velocity standards, BM for benchmark stars, or CR for
CoRoT targets; {\em (3)} the field name; {\em (4)} and {\em (5)} the 2MASS J and
K magnitudes, when available.}
\label{tab:cnames}      
\centering          
\begin{tabular}{l l l l r}     
\hline\hline       
CNAME           & Type & Field    & J     & K \\ 
                &      &          & (mag) & (mag) \\ 
\hline 
19250371+0049014 & CR & Corot     & 11.27 & 10.47 \\
21295872+1208321 & GC & M15       & 11.78 & 11.39 \\
11091266-5837236 & OC & NGC3532   & 11.63 & 11.32 \\
07391749+0513163 & BM & Procyon   & -9.99 & -9.99 \\
ssssssss-sssssss & BM & Sun       & -9.99 & -9.99 \\
21534196-2840169 & RV & HIP108065 & -9.99 & -9.99 \\
\hline \hline
\end{tabular}
\end{table}

To minimize the impact on the total observing time assigned by ESO, calibration
observations are carried out as much as possible in twilight. This is
generally appropriate for the brightest objects. Wavelength calibration lamps are
switched on during GIRAFFE observations for RV standards, while the usual GES
procedure of inserting short exposures with the lamps on is employed for
benchmarks and cluster observations, to avoid spoiling scientific exposures with
scattered light from lamps.

All calibration data are reduced in the same way as any other GES observation to
extract the final science-ready spectra. The ESO processing pipelines
\citep{uves} are employed to produce extracted and wavelength-calibrated UVES
spectra, while a dedicated pipeline for the GIRAFFE processing was developed at
CASU\footnote{http://www.ast.cam.ac.uk/~mike/casu/} (Cambridge Astronomy Survey
Unit). Both pipelines are complemented with GES-specific software to perform
additional operations like sky subtraction or continuum normalization, radial
velocity determination, and so on \citep[for more details, see][for GIRAFFE and
UVES, respectively]{jeffries14,sacco14}.


\section{Basic calibrators}
\label{sec-basic}

Basic spectroscopic calibrations generally include --- besides the acquisition of
an adequate set of calibration frames like bias, flat fields, wavelength
calibration lamps, sky fibers placement\footnote{For young clusters or objects
where the sky is expected to vary significantly across the FLAMES field of view,
the sky fibers positioning and sky subtraction method are crucial.}, and the like
--- the observation of {\em flux standard stars}, {\em radial velocity standard
stars}, and hot, fast rotating stars for telluric absorption band removal, also
referred to as {\em telluric standard stars}. 

\begin{table}
\caption{Radial Velocity standards for zero-point calibration of GES, with their
references RV measurements, taken from \citet{soubiran13} except for GJ~388
\citep{chubak12}.}            
\label{tab-rv}      
\centering          
\begin{tabular}{l l r r l r}     
\hline\hline       
ID         & Type       & V      &  RV      & $\delta$RV \\ 
           &            & (mag)  &  (km/s)  & (km/s) \\ 
\hline   
GJ 388     & M4.5       &  9.43  &   12.453 & 0.066 \\
HIP~616    & K0V        &  8.70  & --42.994 & 0.009 \\
HIP~5176   & G0         &  8.15  &   10.366 & 0.006 \\
HIP~85295  & K7V        &  7.54  & --23.422 & 0.016 \\
HIP~17147  & F9V        &  6.68  &  120.400 & 0.007 \\
HIP~20616  & G0         &  8.41  &   38.588 & 0.009 \\
HIP~26335  & K7         &  8.78  &   21.772 & 0.006 \\
HIP~26973  & K0V        &  8.52  &   26.600 & 0.006 \\
HIP~29295  & M1/M2V     &  8.15  &    4.892 & 0.009 \\
HIP~31415  & F6V        &  7.70  &  --7.479 & 0.012 \\
HIP~32045  & K5         &  8.49  &   40.722 & 0.007 \\
HIP~32103  & G5/G6IV/V  &  8.53  &   27.167 & 0.006 \\
HIP~33582  & G0         &  9.02  & --94.239 & 0.006 \\
HIP~38747  & G5         &  8.37  &  --8.002 & 0.007 \\
HIP~45283  & G2V        &  8.01  &   39.451 & 0.005 \\
HIP~47513  & M2         & 10.38  &   11.626 & 0.007 \\
HIP~47681  & G5V        &  8.41  &   11.289 & 0.007 \\
HIP~50139  & G1V        &  7.75  & --21.976 & 0.005 \\
HIP~51007  & M0         & 10.15  &   21.758 & 0.006 \\
HIP~58345  & K4V        &  6.99  &   48.605 & 0.009 \\
HIP~65859  & M1V        &  9.05  &   14.386 & 0.009 \\
HIP~66032  & K2IV/Vp... &  9.17  &    4.126 & 0.009 \\
HIP~77348  & G5         &  8.05  &    1.907 & 0.011 \\
HIP~80423  & G3/G5Vw... &  9.32  & --42.148 & 0.006 \\
HIP~85295  & K7V        &  7.54  & --23.422 & 0.016 \\
HIP~93373  & G8V        &  8.60  & --91.911 & 0.006 \\
HIP~104318 & G5         &  8.01  &    4.910 & 0.006 \\
HIP~105439 & K0 III+... &  6.75  &   17.322 & 0.006 \\
HIP~106147 & K4/K5V     &  9.11  & --84.533 & 0.009 \\
HIP~108065 & K0/K1III+. &  7.82  & --41.660 & 0.010 \\
HIP~113576 & K5/M0V     &  7.88  &   16.138 & 0.010 \\
\hline                                                
\end{tabular}                                         
\end{table}

For a large spectroscopic survey like GES, where the main deliverables are
chemical abundances, RVs, and APs, the flux calibration of spectra is not a
crucial requirement and thus it is not performed. The correction for telluric
absorption features is likewise not crucial, especially because it only affects
the very last portion of HR21 GIRAFFE spectra, and short wavelength intervals in
the UVES spectra\footnote{Some key diagnostics, like the forbidden oxygen line at
6300~\AA, are indeed affected by telluric absorption, and therefore we anticipate
that a correction for telluric bands will be necessary for a detailed study of
those diagnostics.}. If it will become necessary for specific scientific
applications, telluric absorption bands can be efficiently removed, in future GES
releases, with the use of Earth atmospheric models \citep[for example, from the
TAPAS collaboration,][]{tapas}. Therefore, no observations of telluric standards
were carried out over the current survey, and none are overall planned.

Accurate and precise RV measurements are one of the main tools to fulfil the
scientific goals of GES, and thus a specific calibration strategy was implemented.

\subsection{Radial velocity standard stars}
\label{sec-rv}

GES requires radial velocities with a precision in the range 0.3--1.0
km~s$^{-1}$ to fulfil its various scientific goals \citep{gilmore12}, and both
UVES and GIRAFFE have the potential of delivering RVs with a precision well
below 500~m~s$^{-1}$ \citep[see also][]{sacco14,jackson15}. To reach an accuracy
comparable to the quoted precision, it is necessary to keep the systematics
under control, especially those related to the wavelength calibration scale, the
non-uniform fiber/slit illumination, and the template mismatch in the
cross-correlation procedure. 

For GIRAFFE, to greatly reduce the systematics associated with the wavelength
calibration, it was sufficient to associate to the scientific exposures short
adjacent exposures with the SIMCAL (the simultaneous wavelength calibration lamp)
switched on. Also the use of sky-lines can improve the RV accuracy, as shown by
\citet{jeffries06} and \citet{koposov11}.  To reach an even better accuracy,
below $\simeq$300~m~s$^{-1}$, the repeated observation of RV standards of
different spectral types with the specific purpose of calibrating the RV zero
point is necessary. For UVES, the use of sky lines has proved to reach a
sufficient zero-point accuracy, thus no more UVES observations of RV standards
were required starting from 2015, while for GIRAFFE they are continuing. More
details on the wavelength and RV calibration strategy for GIRAFFE and UVES
spectra, respectively, can be found in \citet{jeffries14}, \citet{sacco14},
and in the GES description papers (Gilmore et al.; Randich et al., in
preparation).

   \begin{figure}
   \centering
   \includegraphics[width=\hsize]{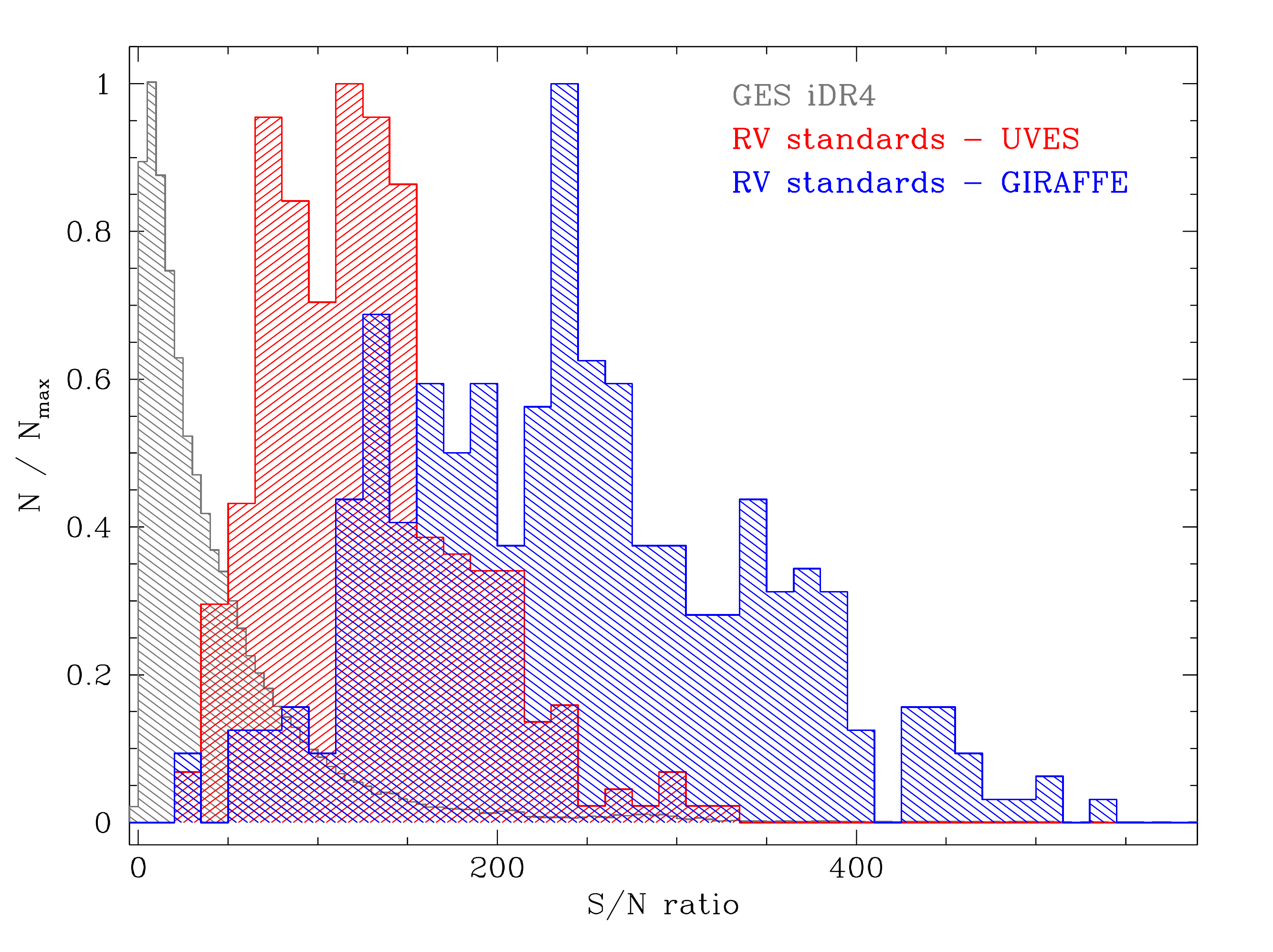}
      \caption{S/N ratios of individual spectra of RV standards for the RV zero
      point calibration. There are between 2 and 20 spectra per star, typically
      10. The UVES setup used is 580 (see Table~\ref{tab-setup}) and the GIRAFFE
      setups are HR9B, HR10, 15N, and 21. The very high S/N ratios are due to
      the RV standards brightness (see Table~\ref{tab-rv}) and to the need of
      integrating for relatively long exposure times, to average out
      illumination non-homogeneities within the fibres.} 
          \label{fig-rvsnr}
   \end{figure}

   \begin{figure}
   \centering
   \includegraphics[width=\hsize]{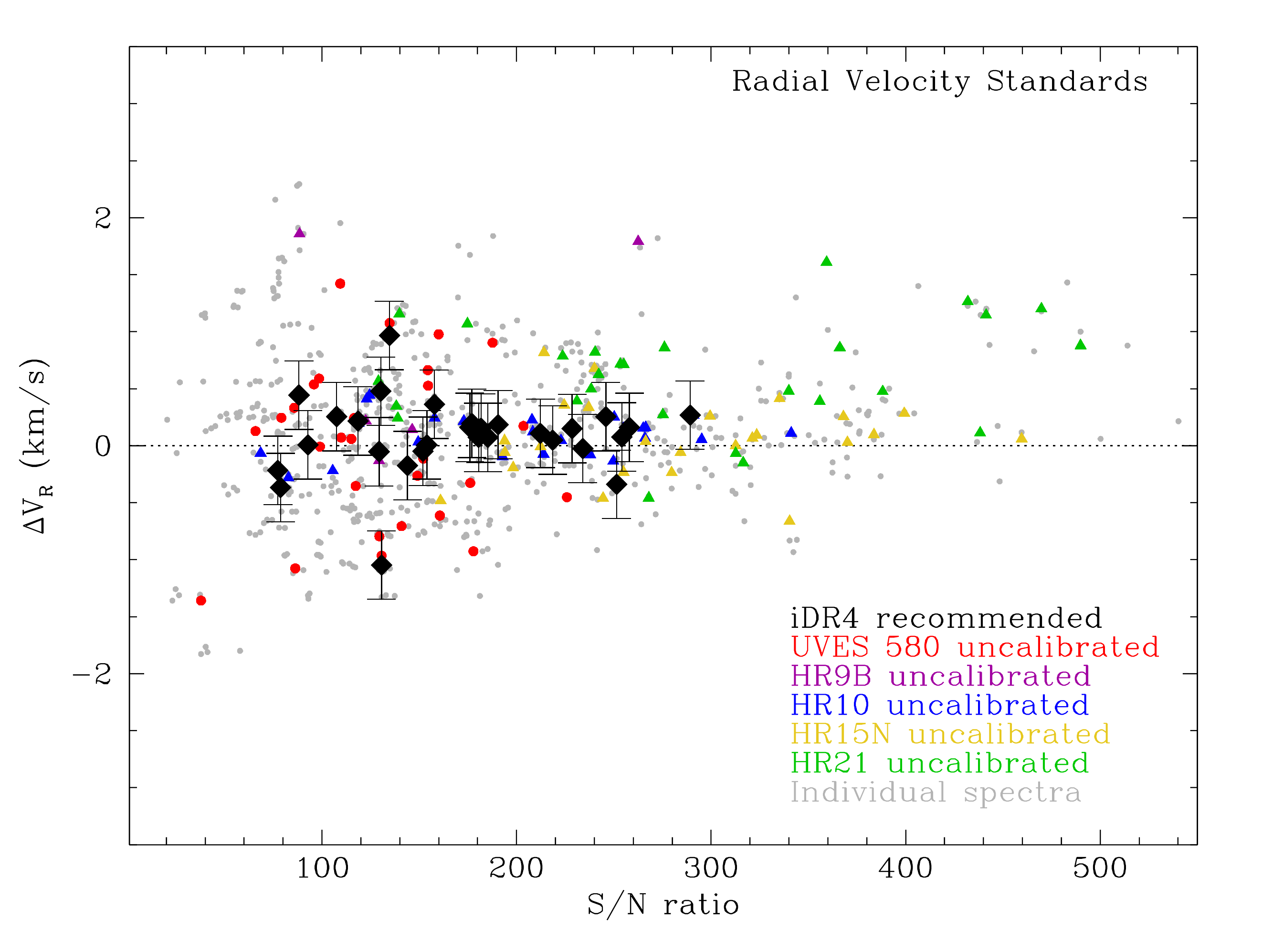}
      \caption{Example of the result of RV homogenization on RV standards. Grey
      dots show the difference between individual spectra measurements and the
      reference values of Table~\ref{tab-rv}. The coloured symbols are the same
      RV differences, but aggregated for each RV standard star in the various
      setups, and are still uncalibrated. The final iDR4 recommended values,
      obtained from the internal homogenization process, are shown by large
      black diamonds, which are placed at the average S/N ratio of the spectra
      obtained for each RV standard star. }  
         \label{fig-rvres}
   \end{figure}

GES was conceived to achieve its maximum impact once combined with Gaia data
(Section~\ref{sec-intro}), therefore the main source of RV standards for GES was
the Gaia standard stars catalogue \citep{soubiran13}, complemented by
\citet{chubak12}. We relied on the best RV calibrators found in the Gaia
catalogue, that appeared to be stable in RV within a few m~s$^{-1}$ over the
explored time baseline (see Table~\ref{tab-rv} for a list of targets). Later,
after the processing of the first internal data release (iDR1), the need for more
RV stars cooler than $\simeq$4000~K emerged, and four M stars were included into
the list. No hot standards are included in the calibration set. We are observing
one or two RV standards in every observing run (approximately once per month). We
used relatively long exposure times (about 100~s) compared to other bright
calibrators like benchmark stars, avoiding saturation, not only to increase the
S/N ratio (see Figure~\ref{fig-rvsnr}), but also to ensure uniform slit
illumination, and with the SIMCAL on when observing with GIRAFFE. 

Besides being used to set the zero-point of GIRAFFE RV measurements, the RV
standards are also used in the WG15 homogenization procedure, described in detail
by H17. Briefly, the performance of each of the observed setups was tested with
RV standards (see Figure~\ref{fig-rvres}), to identify the setups that show the
smallest offset with respect to the reference values of Table~\ref{tab-rv}. All
other setups were corrected to the scale of the best setup (generally HR10,
followed by HR15N) using the stars in common with the best available setup to
compute an offset. In previous GES releases, offsets of $\simeq$0.5~km~s$^{-1}$
were reported between UVES and GIRAFFE \citep[see][among
others]{sacco14,donati14,lardo15}. In iDR4, the two instruments showed much
smaller differences, of a few meters per second, thanks to the use of sky lines
to correct for UVES wavelength calibration uncertainties. The setup that shows
the largest offset is HR21 ($\simeq$0.5~km~s$^{-1}$), for which the SIMCAL lamps
are switched off to avoid contaminating the scientific exposure given the high
efficiency of this particular setup. 


\section{Benchmark stars}
\label{sec-bench}

   \begin{figure}
   \centering
   \includegraphics[width=\hsize]{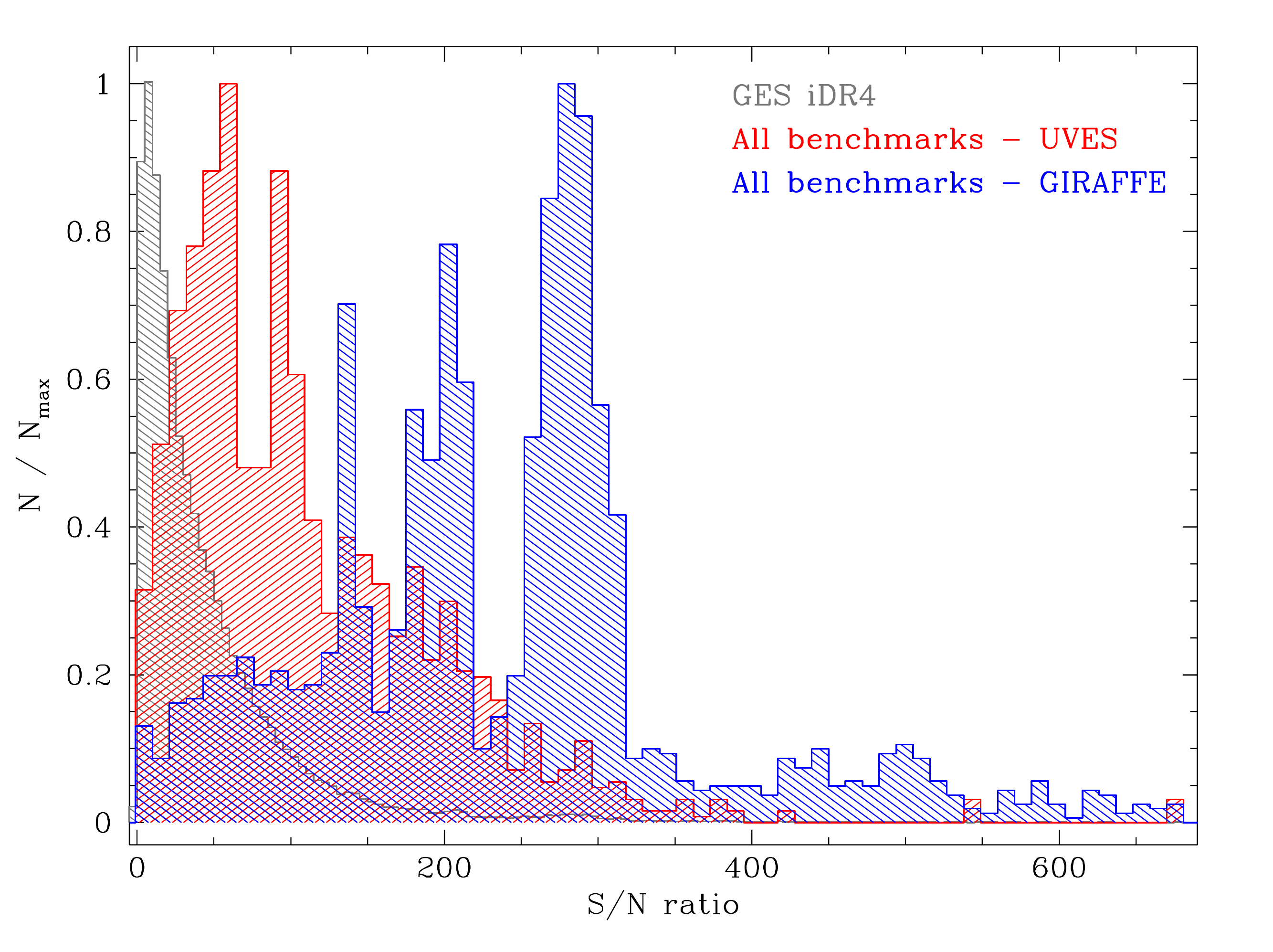} \caption{S/N ratio of individual
      spectra of benchmark stars. In an initial phase, spectra were obtained in
      a range of S/N ratio values. Later, we aimed at obtaining at least 3
      exposures per benchmarks star per setup, with a combined S/N$>$100 per
      pixel (without saturating).}
      \label{fig-bsnr}
   \end{figure}
\begin{table}
\caption{Gaia FGK benchmark stars observed in GES. Magnitudes and APs are from
         \citet{heiter15b}, NLTE-corrected metallicities from
         \citet{jofre14}.}            
\label{tab-bs}      
\centering          
\begin{tabular}{l@{ }c c c c c}     
\hline\hline       
ID             & Type     & V      & [Fe/H]$_{\rm{NLTE}}$ &T$_{\rm{eff}}$& log$g$       \\ 
               &          & (mag)  & (dex)  & (K)          & (dex)        \\ 
\hline      
Procyon        & F5IV-V   &  0.366 &  +0.01 & 6545 & 4.00 \\
HD 84937       & sdF5     &  8.324 & --2.03 & 6275 & 4.06 \\
HD 49933       & F2V      &  5.762 & --0.41 & 6635 & 4.21 \\
$\delta$ Eri   & K1III-IV &  3.527 &  +0.06 & 5045 & 3.76 \\
HD 140283      & sdF3     &  7.210 & --2.36 & 5720 & 3.67 \\
$\epsilon$ For & K2V      &  5.883 & --0.60 & 5069 & 3.45 \\
$\eta$ Boo     & G0IV     &  2.681 &  +0.32 & 6105 & 3.79 \\
$\beta$ Hyi    & G0V      &  2.797 & --0.04 & 5873 & 3.98 \\
$\alpha$ Cen A & G2V      &  0.002 &  +0.26 & 5847 & 4.31 \\
HD 22879       & F9V      &  6.689 & --0.86 & 5786 & 4.23 \\
Sun            & G2V      &--26.74 &   0.00 & 5771 & 4.44 \\
$\tau$ Cet     & G8.5V    &  3.495 & --0.49 & 5331 & 4.44 \\
$\alpha$ Cen B & K1V      &  1.357 &  +0.22 & 5260 & 4.54 \\
18 Sco         & G2Va     &  5.505 &  +0.03 & 5747 & 4.43 \\
$\mu$ Ara      & G3IV-V   &  5.131 &  +0.35 & 5845 & 4.27 \\
$\beta$ Vir    & F9V      &  3.608 &  +0.24 & 6083 & 4.08 \\
Arcturus       & K1.5III  &--0.051 & --0.52 & 4247 & 1.60 \\
HD 122563      & F8IV     &  6.200 & --2.64 & 4587 & 1.61 \\
$\epsilon$ Vir & G8III    &  2.828 &  +0.15 & 4983 & 2.77 \\
$\xi$ Hya      & G7III    &  3.541 &  +0.16 & 5044 & 2.87 \\
$\alpha$ Tau   & K5III    &  0.867 & --0.37 & 3927 & 1.22 \\
$\psi$ Phe     & M4III    &  4.404 & --1.24 & 3472 & 0.62 \\
$\gamma$ Sge   & M0III    &  3.476 & --0.17 & 3807 & 1.05 \\
$\alpha$ Cet   & M1.5IIIa &  2.526 & --0.45 & 3796 & 0.91 \\
$\beta$ Ara$^a$ & K3Ib-II &  2.842 & --0.05 & 4197 & 1.05 \\
HD 220009$^a$  & K2III    &  5.047 & --0.74 & 4217 & 1.43 \\
HD 107328      & K0IIIb   &  4.970 & --0.33 & 4496 & 2.09 \\
$\epsilon$ Eri & K2Vk:    &  3.726 & --0.09 & 5050 & 4.60 \\
\hline                                                
\multicolumn{6}{l}{$^{a}$Not recommended as benchmarks from iDR5 on.}\\
\end{tabular}                                         
\end{table}

In traditional works of stellar abundance analysis, the Sun is used as a
reference, either to verify a posteriori the validity of the presented results
by performing an analysis of a Solar spectrum with the same technique employed
on the programme stars or to perform a differential analysis of the programme
stars with respect to the Sun \citep[see][for a GES-related example of this type
of analysis]{sousa14}. A second example of a reference star widely used for
testing abundance analysis of cooler, more metal-poor stars is Arcturus
\citep[see, e.g.,][and included references]{ramirez11,meszaros13,morel14b}.
Moreover, when large samples are analyzed, the stars in common among different
literature studies are used as a comparison to put all data on the same system,
as much as possible \citep[see, e.g.,][]{worley12,depascale14,bensby14}. Within
the Gaia mission preparatory effort, the concept of one reference star for APs
determination and abundance analysis verification has been extended to define
the so-called {\em Gaia benchmark stars} set \citep{heiter15b}. Benchmark stars
ideally have known Hipparcos parallaxes, angular diameters, and bolometric
fluxes, and their masses have been determined in a homogeneous way, so their
effective temperatures and surface gravities can be derived as independently as
possible from spectroscopy.

Even if FLAMES is not the ideal instrument to observe individual stars, it was
deemed extremely important to observe these fundamental reference objects within
GES. Being bright stars, they were observed mainly during twilight, with the
three GES UVES setups and with the GIRAFFE HR9B, 10, 15N, and 21 setups (see
Table~\ref{tab-setup}), i.e., the four GES setups used for FGK stars in the MW
field and in OCs. GES further extended the list to include also a few cooler K
and M benchmarks and a few hotter O, B, and A benchmarks, as detailed in the
next sections. The hot benchmark stars were observed also with the GES hot
stars GIRAFFE setups: HR3, 5A, 6, and 14A (see Table~\ref{tab-setup}). The
S/N ratio of the observed spectra is reported in Figure~\ref{fig-bsnr}. 

   \begin{figure}
   \centering
   \includegraphics[width=\hsize]{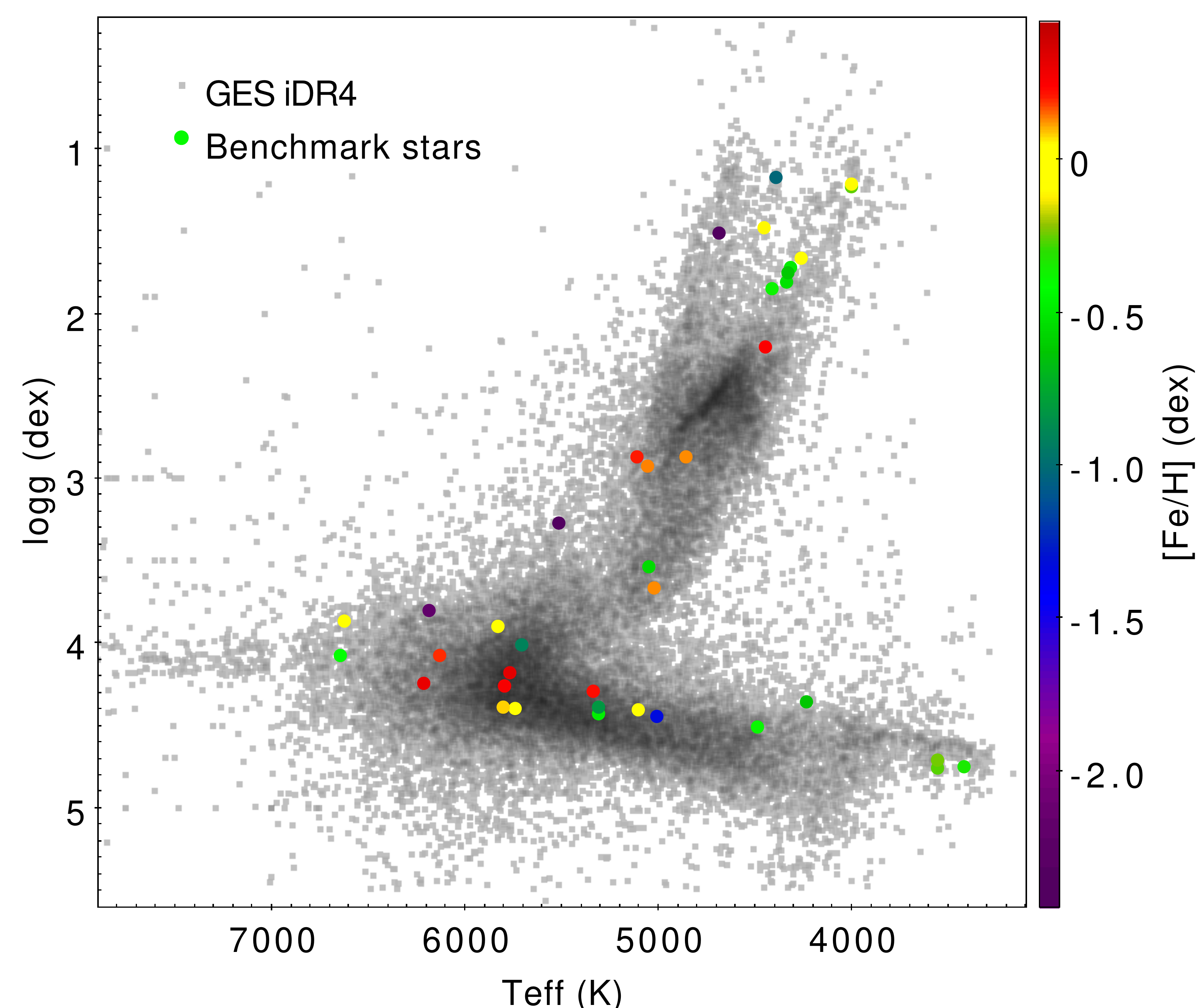} \caption{Position on the
      T$_{\rm{eff}}$-log$g$ plane of the Gaia FGK benchmark stars \citep[see also
      Section~\ref{sec-benchmarks}]{heiter15b} analyzed in iDR4, coloured according
      to their [Fe/H]. A few of the cooler benchmarks (T$_{\rm{eff}}<$4000~K)
      described in Section~\ref{sec-cool} were also analyzed in iDR4. The whole
      GES iDR4 sample is reported in the background as smaller grey squares.}
      \label{fig-bs}
   \end{figure}

Benchmark stars and candidate benchmarks are used both in the WG-level and
survey-level homogenization processes, to assess which abundance analysis nodes
and WGs, respectively, perform better in different regions of the parameters
space, as expanded in Section 4.4. More details on the use of benchmark stars
can be found in \citet{smiljanic14}, \citet{lanzafame15}, and H17.


\subsection{Gaia FGK benchmarks} 
\label{sec-benchmarks}

The FGK benchmark stars that were selected as GES astrophysical calibrators are
listed in Table~\ref{tab-bs}. They are extracted from the original set of Gaia
FGK benchmark stars \citep{heiter15b}, which contains 34 stars with
T$_{\rm{eff}}$ in the range $\simeq$3500--6500~K, log$g$ in
$\simeq$0.5--4.5~dex, and with a metallicity ranging from super-solar to
--2.5~dex. Additional spectra of these stars were gathered from the ESO archive
(UVES and HARPS) and from the NARVAL archival observations at the Pic du Midi,
and homogenized into a comprehensive library of high-resolution spectra 
\citep{blanco14}. The only fundamental parameter that was not well constrained
for these stars in the literature was [Fe/H], thus an effort from the GES
abundance analysis nodes was made to derive independently a set of reference
[Fe/H] value for each of them \citep{jofre14,jofre15}, along with abundances for
10 elements, as a first step. Figure~\ref{fig-bs} gives an idea of the parameter
space covered by the Gaia benchmark stars. The set only contains few metal-poor
stars, a regime that is not much sampled in GES (see Figure~\ref{fig-gesmet}),
but we recently identified a few more candidates with [Fe/H]$<$--1.2~dex
\citep{hawkins16}. 


\subsection{Additional M benchmarks}
\label{sec-cool}

The collection of Gaia benchmark stars, from which we selected the sample
described in the previous section, does not include a sufficient number of stars
cooler than $\simeq$3500~K. Benchmark stars in the M-dwarf region are needed
both for Gaia (expected to observe more than one million M dwarfs) and GES,
where M dwarfs are included among OC target stars. Angular diameter measurements
for potential cool benchmark stars have only recently started to become
available, and homogeneous metallicity determinations for the most promising
ones are not available yet. 

\begin{table}
\caption{Additional M-dwarf benchmark stars, with their magnitudes and spectral
type from SIMBAD, metallicities as noted, and T$_{\rm{eff}}$ from
\citet{boyajian12}.}            
\label{tab-cool}      
\centering          
\begin{tabular}{l l c c c l}     
\hline\hline       
ID          & Type        & V      & [Fe/H] & T$_{\rm{eff}}$ & Status \\ 
            &             & (mag)  & (dex)  & (K)           \\ 
\hline 
GJ 205 & M1.5V       &  7.97  &  +0.35$^a$ & 3801 & iDR3 \\
GJ 436 & M3V         & 10.59  &  +0.03$^b$ & 3416 & iDR4 \\
GJ 526 & M1.5V       &  8.50  & --0.30$^a$ & 3618 & iDR4 \\
GJ 551$^c$ & M5.5V   & 11.05  &  +0.24$^d$ & 3054 & --- \\
GJ 581 & M2.5V       & 10.61  & --0.02$^b$ & 3442 & iDR4 \\
GJ 699 & M4V         &  9.51  & --0.39$^a$ & 3224 & iDR3 \\
GJ 880 & M1.5V       &  8.64  &  +0.03$^e$ & 3713 & iDR2 \\
\hline                                                  
\multicolumn{6}{l}{$^{a}$Metallicity from \citet{rojas12}.}\\
\multicolumn{6}{l}{$^{b}$Metallicity from \citet{lindgren16}.}\\
\multicolumn{6}{l}{$^{c}$Not observed yet, we will rely on UVES archival data.}\\
\multicolumn{6}{l}{$^{d}$Metallicity from \citet{jofre14}, considering
$\alpha$~Cen A and B.}\\
\multicolumn{6}{l}{$^{e}$Metallicity from \citet{neves14}.}\\
\end{tabular}                                         
\end{table}

Nevertheless, we selected a number of candidate benchmarks among the best
studied M-dwarf stars, listed in Table~\ref{tab-cool}. For four of these stars,
angular diameters were published by \citet{boyajian12} with a precision of
1--2\%, while the angular diameters of GJ~436 and GJ~581 were determined by
\citet{vonbraun11,vonbraun12} to 3\%, and that of GJ~551 by \citet{demory09} to
5\%. Bolometric fluxes were measured for all stars by \citet{boyajian12} with a
precision of 1\%. These data give T$_{\rm{eff}}$ independent from photometry or
spectroscopy for all stars, as listed in Table~\ref{tab-cool}. Spectroscopic
metallicity determination is more difficult for M dwarfs than for FGK dwarfs due
to the more complex optical spectra. Several approaches have been pursued in the
literature. These include calibrations of photometric data or low-resolution
infrared spectroscopic features \citep[e.g.,][]{rojas12}, or analysis of
high-resolution spectra in optical or infrared regions
\citep[e.g.,][]{onehag12,neves14,lindgren16}. Usually, samples of binaries with
M- and FGK components are used for calibration or validation of the methods.
Selected metallicities from various sources are listed in Table~\ref{tab-cool}.

For most of these stars, there are additional high-resolution spectra available,
in addition to those obtained with the GES setups. GJ~436, GJ~526, and GJ~880
were observed at optical and near-IR wavelengths with the NARVAL spectrograph.
For GJ~436, GJ~551, GJ~581, and GJ~880, J-band spectra with R=$50\,000$ were
obtained with the CRIRES spectrograph at the VLT. GJ~699 is included in the
CRIRES-POP library \citep[wavelength range from 1 to 5~$\mu$m,][]{lebzelter12}.
These high-quality archival data constitute a legacy sample that will allow us to
compare results obtained in the optical and infrared wavelength regions. In GES
iDR4 all observations for the listed cool benchmarks were completed, except for
GJ~551 (Proxima Centauri), for which we will most probably have to rely on UVES
archival data in future data releases. 


\subsection{Additional OBA benchmarks}
\label{sec-warm}

While benchmarks stars -- with APs as independent as possible from spectroscopy --
are becoming available for FGK and M types, as we discussed above, the situation
is not as favourable in the case of hotter stars. This is due to the lack of
interferometric data and the lack of spectrophotometry in the ultraviolet where
the flux of these stars dominates. With this limitation in  mind, one can,
however, define a sample of well-studied A, B, and O-type stars  with relatively
well-established parameters in the refereed literature -- even if not independent
from spectroscopy. 

\begin{table}
\caption{List of OBA benchmark candidates observed by GES, none was analyzed in
any internal release so far. A few more OBA stars will be observed before the end
of the survey. Magnitudes and spectral types are from SIMBAD.} 
\label{tab-warm}
\centering
\begin{tabular}{l@{}l@{}l@{}lll}
\hline\hline
Star            &  V    &       Type &T$_{\rm eff}$&$\log g$& Status \\
                & (mag)~~ &            & (K)       &  (dex) & \\
\hline
HD 93128$^a$    &  6.90 & O3.5V      &  49\,300  &  4.10  & started \\
HD 319699$^a$   &  9.63 & O5V        &  41\,200  &  3.91  & started \\
HD 163758$^a$   &  7.32 & O6.5Iafp~~ &  34\,600  &  3.28  & observed \\
HD 68450$^a$    &  6.44 & O9.7II     &  30\,600  &  3.30  & observed \\
$\tau$ Sco$^{b,c,d,e,f,g}$~~~     
                &  2.81 & B0.2V      &  31\,750  &  4.13  & observed \\
$\theta$ Car$^{c,d}$
                &  2.76 & B0Vp       &  31\,000  &  4.20  & observed \\
$\gamma$ Peg$^{b,c,h}$    
                &  2.84 & B2IV       &  22\,350  &  3.82  & observed \\
HD 56613$^c$    &  7.21 & B8V        &  13\,000  &  3.92  & observed \\
134 Tau$^i$     &  4.87 & B9IV       &  10\,850  &  4.10  & observed \\
68 Tau$^j$      &  4.31 & A2IV       &   9\,000  &  4.00  & observed \\
\hline \hline
\end{tabular}
\begin{flushleft} 
References for APs: (a) Holgado et al., in preparation (see text); 
(b) \citet{nieva12}; (c) \citet{lefever10}; (d) \citet{hubrig08}; 
(e) \citet{simondiaz06}; (f) \citet{mokiem05}; (g) \citet{martins12}; 
(h) \citet{morel08a}; (i) \citet{smith93}; (j) \citet{burkhart89}.
\end{flushleft}
\end{table}

For the calibration of APs of A-type stars, we selected 5 benchmark stars
previously observed for the AP calibration of hot stars for Gaia
\citep{bailerjones13}. These Gaia benchmark stars were observed with
S/N$\approx$1000, using the Hermes@Mercator (R=85\,000) in La Palma, Spain.
Additional Gaia OBA benchmark spectra are being observed in ongoing dedicated
observing programs. We complemented the set with one late B-type star with
T$_{\rm eff}$$\approx$11\,000~K (134~Tau). These stars were carefully selected
to cover different spectral subtypes, to have small $v \sin i$ values, and to be
bright and visible from Paranal. Their optical spectra show sufficiently deep
and narrow absorption lines in the wavelength regions also observed by GES. They
are presently being observed by GES and will be used not only for survey-level
homogenization, but also for testing the quality of APs and elemental abundance
computed by the WG13 nodes, for all GES A- and late B-type stars in various
Galactic young OCs. 

For the early B-type stars, the selected pool of candidate benchmarks had their
parameters (T$_{\rm eff}$ and log $g$) estimated solely from high-resolution
spectroscopic data (e.g., this excludes T$_{\rm eff}$ measurements based on
photometric indices). Also, only studies treating the line formation in non-LTE
were considered. The model atmospheres used may be either LTE or non-LTE
\citep[LTE being a reasonable assumption for B-type dwarfs;][]{przybilla11}, but a
full line blanketing was considered a requirement. We performed a comparison of
the available studies for each candidate B-type benchmark for GES, and rejected
discrepant measurements \citep[e.g., a few of the very high gravities from][and
references therein]{daflon04}. In some cases, stars were studied by various
authors with similar data and methods, but we preferred one set over another to
avoid redundancies. For example we used the results of \citet{nieva12} for the
four stars in common with \citet{nieva11}, or for the three stars in common with
\citet{irrgang14}. The stars eventually selected have consistent APs from at least
two high-quality and independent studies. It is important to note that most B stars analysed by GES, and
generally belonging to young OCs, are fast rotators (e.g., $\langle v\sin
i\rangle$ $\sim$ 160 km s$^{-1}$ in NGC 3293). On the contrary, the abundance
studies in the literature are heavily biased against such objects. As a
consequence, the vast majority of the selected B benchmark stars are slow rotators
\citep[by far the fastest rotator is $\theta$ Car with  $v\sin i \sim$ 110 km
s$^{-1}$;][]{hubrig08}. This caveat should be kept in mind. 

   \begin{figure}
   \centering
   \includegraphics[width=\hsize]{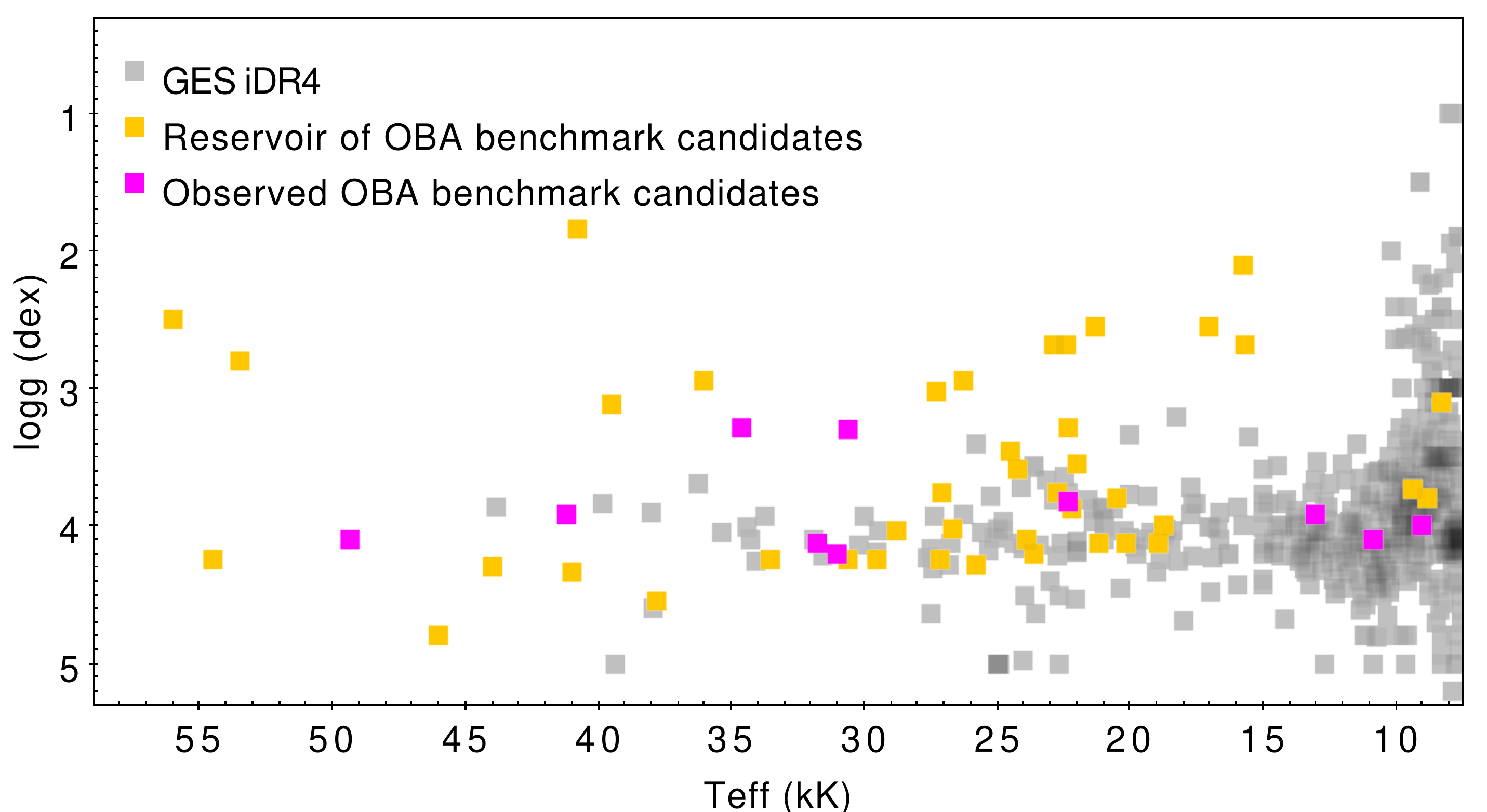}
      \caption{Position of OBA benchmark candidates on the T$_{\rm{eff}}$-log$g$
      plane. The OBA benchmarks are plotted as magenta squares; the pool of OBA
      benchmarks from which a few more will be selected for observations is
      represented by yellow squares; the GES iDR4 sample is reported in the
      background as grey squares.} 
      \label{fig-warm}
   \end{figure}

The O-type candidate benchmarks were selected from the new Galactic O-Star
Spectroscopic Survey spectral classification standard grid \citep{maiz15}, which
is a recent revision of the atlas for spectral classification, first established
by \citet{walborn90}. The full grid comprises more than 100 stars with spectral
subtypes from O2 to O9.7 and luminosity classes from V to Ia, in both hemispheres,
and it has been observed at high resolution (R$\simeq$50\,000) in two dedicated
surveys \citep[OWN and IACOB, see][]{barba10,barba14,simon11b,simon11c,simon15}. A
quantitative and homogeneous spectroscopic analysis of the OWN and IACOB samples
is being performed within the framework of the IACOB project, and the results will
soon be published (Holgado et al., in preparation), along with the full spectra
library. The multi-epoch spectra of the OWN and IACOB projects also allow for
variability searches, and a literature comparison with recent hot star surveys
results for $v\sin i$, T$_{\rm{eff}}$, log$g$, and helium abundance
\citep{repolust04,markova14,martins15}, is also being carried out. 

Table~\ref{tab-warm} lists the OBA candidate benchmarks observed by GES up to
now, while Figure~\ref{fig-warm} shows the parameter coverage of both the
observed and candidate OBA benchmarks in the T$_{\rm{eff}}$-log$g$ plane. We
expect to observe a few more OBA benchmarks before the end of the survey. 


   \begin{figure}
   \centering
   \includegraphics[width=\hsize]{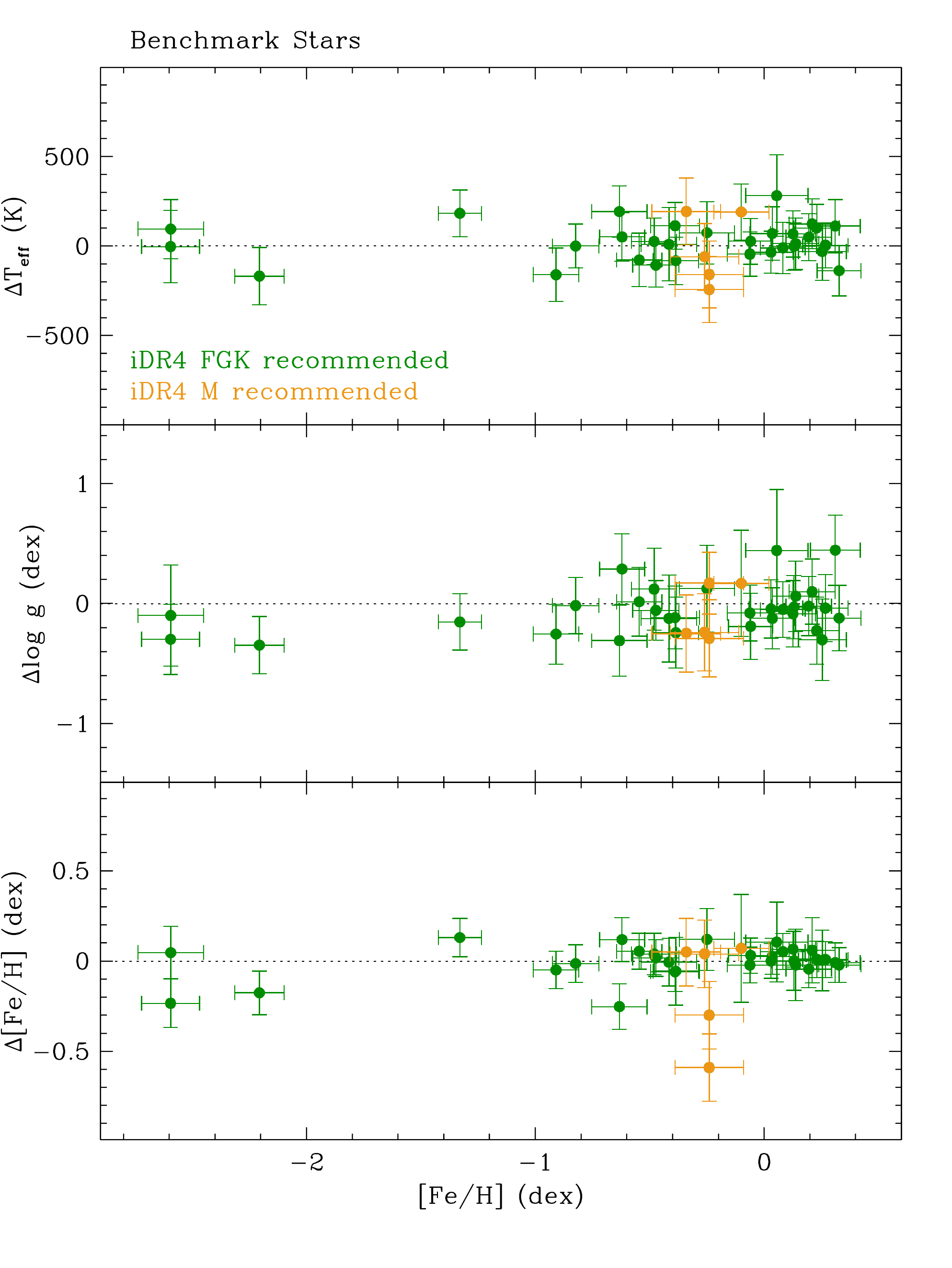}
      \caption{Comparison of GES iDR4 recommended values with the reference APs
      \citep{heiter15b} and [Fe/H] \citep{jofre14} values. In all panels, FGK
      benchmark stars are plotted in green and M benchmarks in orange. All
      differences are in the sense GES minus reference.}
      \label{fig-bsres}
   \end{figure}

\subsection{GES benchmarks results}
\label{sec-bmres}

Benchmark stars were used in GES iDR4 both within each WG to homogenize the
results of different abundance analysis nodes \citep[see, e.g.][]{smiljanic14}
and at the survey level to homogenize the results of different WGs (see H17).
Additionally, in iDR4, the FGK benchmark stars are among the few calibrators that
are also used to provide an external reference for APs, i.e., they are used as
{\em absolute} calibrators. For example in WG11, they are used to define weights
for each of the abundance analysis node results, that vary for different regions
of the AP space based on that node results on benchmark stars. The calibration
procedures derived using benchmarks, among other calibrators, are applied to all
survey data and therefore it is useful to examine the effects of the whole
process on the benchmark stars themselves. Figure~\ref{fig-bsres} shows
differences of the iDR4 recommended GES APs and [Fe/H] values with the reference
AP values \citep{heiter15b} and NLTE metallicities \citep{jofre14}. The average
differences are T$_{\rm{eff}}$=14$\pm$113~K, $\log g$=--0.07$\pm$0.19~dex, and
[Fe/H]=--0.02$\pm$0.13~dex. In all cases, the average offsets are negligible, and
the dispersions give an idea of the typical GES performances on these high S/N
ratio spectra.


\section{Star clusters}
\label{sec-clusters}

   \begin{figure}
   \centering
   \includegraphics[width=\hsize]{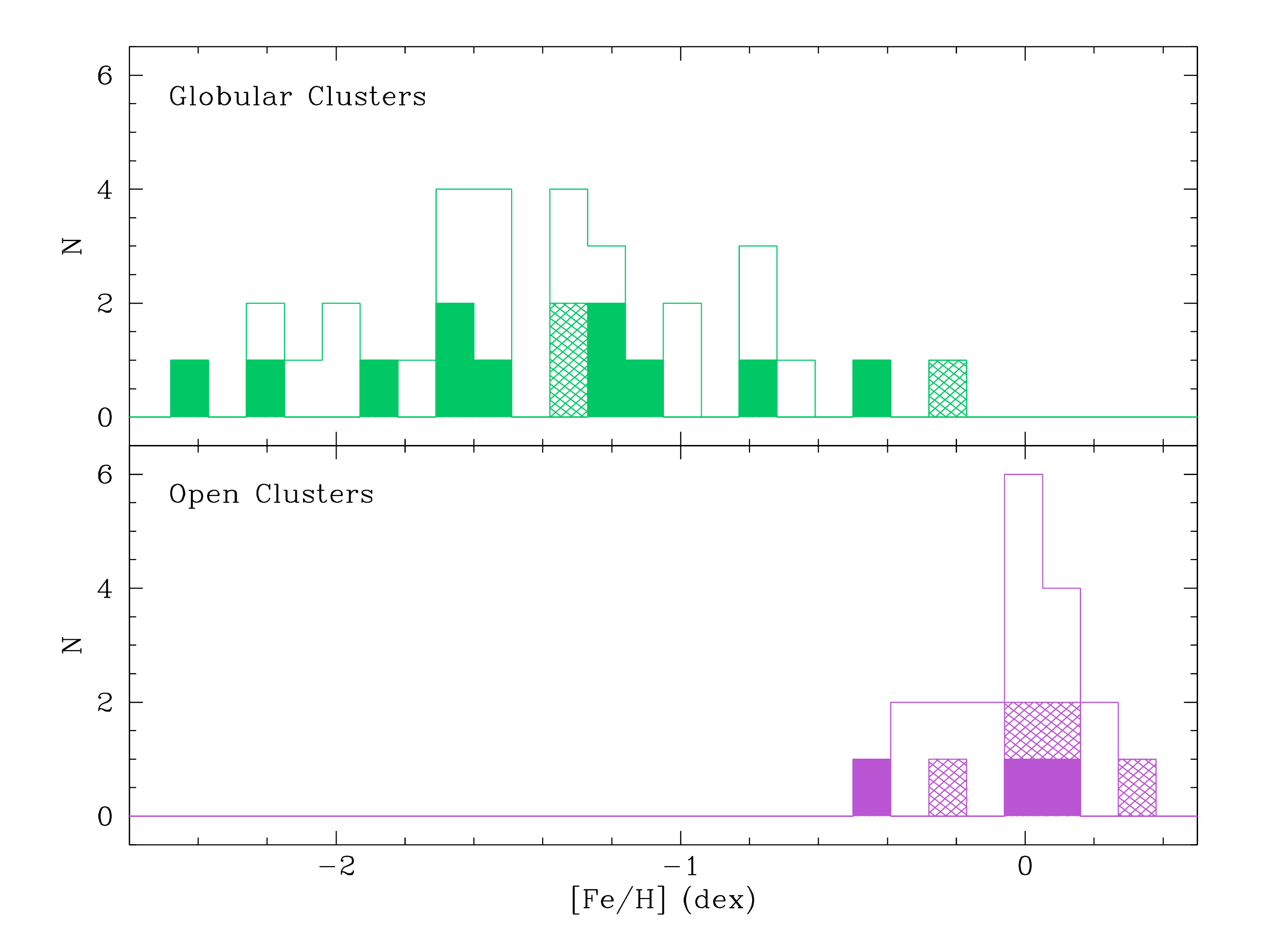}
      \caption{Metallicity distribution of calibrating clusters: the top panel
      shows GCs and the bottom one OCs. Heavily shaded histograms show clusters
      included in iDR4; lightly shaded histograms the ones that will be included
      in future releases (see also Tables~\ref{tab-gc} and \ref{tab-oc}); empty
      histograms represent a pool of viable candidates to complete the coverage.
      A few of them could be selected in the next observing runs, depending on
      observations scheduling and data homogenization needs.}
      \label{fig-gc}
   \end{figure}
   \begin{figure}
   \centering
   \includegraphics[width=\hsize]{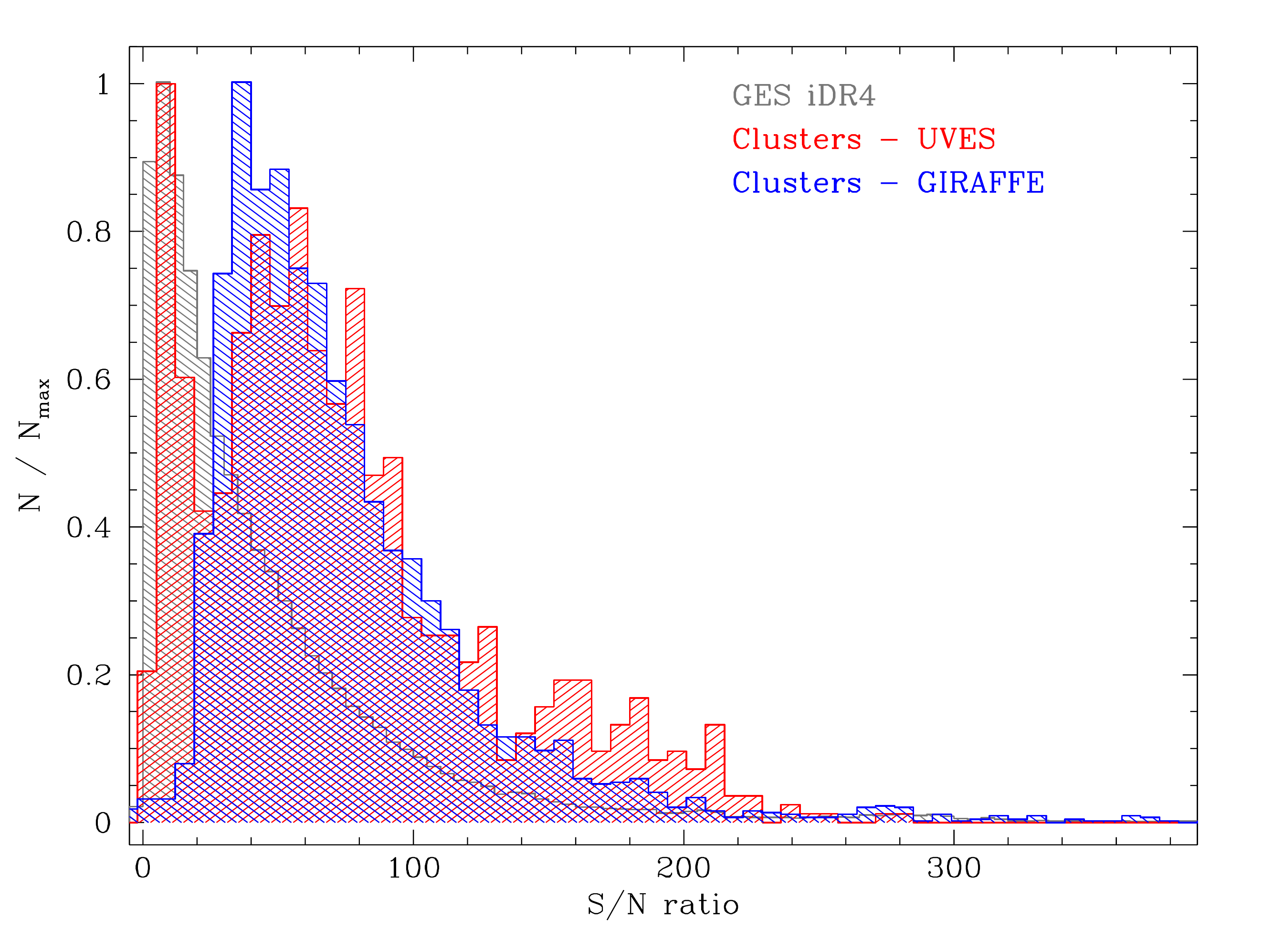}
      \caption{S/N ratio of individual spectra of calibrating cluster stars,
      both in open and globular clusters. There are typically $>$3 exposures per
      star per setup. The required S/N ratio per star --- after combining at
      least three spectra per star --- was $>$50, thus individual exposures peak
      roughly around 30--40 for GIRAFFE, who was driving the total exposure time.
      A tail of low S/N spectra for UVES contains mostly archival spectra of
      subgiants and MS stars.}
         \label{fig-clustersnr}
   \end{figure}

Often, the goal of providing astrophysical calibrations for a spectroscopic
survey is achieved by observing clusters in the MW. Both relatively old OCs
(Section~\ref{sec-open}), and GCs (Section~\ref{sec-globular}) are used in
various surveys (RAVE, GALAH, and APOGEE, for example). They are extremely
powerful calibrators of both APs and abundance ratios, for a number of reasons:

\begin{table*}[!th]
\caption{GES calibrating open clusters, with basic properties from \citet[][and
latest updates]{dias02}, and [Fe/H] metallicity from \citet{heiter14}, except
where noted. The status column refers only to the GES {\em calibration}
observations, i.e., to those OCs that were observed with both the OC observing
setups and the MW ones. The last column indicates other surveys using each OC
as calibrator, along with other useful annotations.}
\label{tab-oc}      
\centering          
\begin{tabular}{l r c c r r c l}     
\hline\hline       
Cluster        & Dist & E(B--V) & Age   & [Fe/H] &$\langle$RV$\rangle$& Status    & Notes \\
               & (pc) & (mag)   & (Gyr) &  (dex) & (km/s)             &           & \\
\hline      
NGC~3532       &  492 &  0.028  &  0.30 &  +0.00 &               4.33 &  in iDR3  & RAVE                     \\ 
NGC~6705 (M11) & 1877 &  0.428  &  0.25 &  +0.12 &              35.08 &  in iDR3  & internal calibrator      \\ 
NGC~2243\tablefootmark{a} & 2450 &  0.060  &  4.00 & --0.48 &   59.84 & in iDR3/4 & APOGEE                   \\ 
Melotte~71     & 3154 &  0.113  &  0.24 & --0.27 &               0.55 & observed  & RAVE, APOGEE             \\ 
NGC~6253       & 1510 &  0.200  &  5.00 &  +0.34 &            --29.40 & observed  & very metal-rich          \\ 
NGC~2420       & 2480 &  0.040  &  2.00 & --0.05 &              73.57 & observed  & APOGEE                   \\ 
NGC~2477       & 1300 &  0.240  &  0.60 &  +0.07 &               7.26 & started  & RAVE, GALAH              \\ 
\hline      
\hline      
\multicolumn{8}{l}{$^{(a)}$Distance, reddening, and age from \citet{bocce1}.}\\
\end{tabular}                                         
\end{table*}

\begin{itemize}
\item{although their APs are not as accurate as those of benchmark stars
(Section~\ref{sec-bench}), clusters contain many stars with similar --- to first
order --- distances, ages, and chemical compositions\footnote{With caution on
some light (C, N, O, Na, Mg) and s-process elements, which can vary
significantly in GC stars \citep[see, e.g.,][]{gratton12}. Also, spreads in
[Fe/H] are observed or claimed in a few GCs. This needs to be duly taken into
account when using these clusters as calibrators.}; thus, clusters provide
extremely robust calibrators, because they also provide a way to statistically
estimate the {\em uncertainty} on determined metallicities and abundance
ratios;}
\item{both OCs and GCs can --- globally --- rely upon a vast literature of
photometric, astrometric, and spectroscopic measurements, and on very advanced
models of stellar structure and evolution, that are invaluable tools, making
clusters ideal reference objects for both {\em external calibration} and {\em
literature cross-checks};}
\item{having stars with virtually the same distance, it is possible to precisely
know surface gravity, which is one of the most difficult quantities to derive for field
stars without an absolute distance determination (see also
Section~\ref{sec-corot}); in general it is possible to derive precise APs from
the many high-quality photometric catalogues available, thus clusters also
provide an invaluable testbench for a survey's APs determination;}
\item{cluster stars have different APs, varying along the sequences of the
colour-magnitude diagram in a regular way,  allowing for the investigation of
chemical abundance trends with parameters: {\em no other calibrator allows for
this kind of check of the internal consistency of an abundance analysis}, which
is invaluable even for each individual method, even before comparing different
methods;}
\item{finally, the AP variations of cluster stars --- having the same
metallicity --- allows for a very efficient {\em internal calibration} of a
complex survey like GES; in particular, they allow the linking of various
abundance analysis techniques employed by the many GES abundance analysis nodes
concerning giants and dwarfs, cool and hot stars, GIRAFFE and UVES spectra.}
\end{itemize}

Star clusters in iDR4 were not used as absolute (external) calibrators like
benchmark stars, but were rather used {\em a posteriori} to verify the quality
of the whole homogenization procedure at the node, WG, and survey levels (see
also Section~\ref{sec-clusterparams}). The metallicity range covered by GES
calibrating clusters is presented in Figure~\ref{fig-gc}, while
Figure~\ref{fig-clustersnr} shows the distribution of S/N ratios for individual
spectra, where typically each star was observed three times per setup. 


\subsection{OC selection criteria}
\label{sec-open}

Calibrating OCs\footnote{We term {\em calibrating OCs} here the OCs (or OC
stars) that are observed specifically for the purpose of calibration, i.e., with
both the MW field setups and the OC setups. Many more OC stars and OCs are
observed for GES scientific purposes, and they will be called {\em science OCs}.
Generally, the calibrating OC stars are a subset of the science OCs.} were
selected to interface with other current spectroscopic surveys, including also
well known and studied clusters, to cover the metallicity range of interest. In
the case of OCs, however, we tried to use as much as possible the targets
selected by the GES OC group, because they already gather state-of-the-art
literature data in terms of photometry, membership, binarity, and so on (see
Bragaglia et al., in preparation, for more details), and because we could profit
from the GES analysis to further select more reliable members. This is also the
reason why calibrating OC observations started later in the survey than GC ones.

We gave priority to relatively old OCs, where a red clump is present\footnote{In
any case, we did not select OCs younger than 100-200~Myrs as calibrators.}, so
that in many calibrating OCs we will have both red clump giants and main
sequence dwarfs. The GES science target OCs are generally observed with the UVES
580 setup, and the GIRAFFE HR15N and HR9B setups (see Table~\ref{tab-setup}).
Additionally, the stars selected for calibrations were also observed with the
HR10 and HR21 GIRAFFE setups, i.e., those used for MW field stars. This was
intended to facilitate the internal calibration and to increase the wavelength
coverage, thus making the abundance analysis of calibrating stars more robust.

In iDR4, three OCs were observed, as indicated in Table~\ref{tab-oc}, while four
more were completed recently. Additional OCs may be added in the future, depending
on scheduling and analysis requirements. 

   \begin{figure}
   \centering
   \includegraphics[width=\hsize]{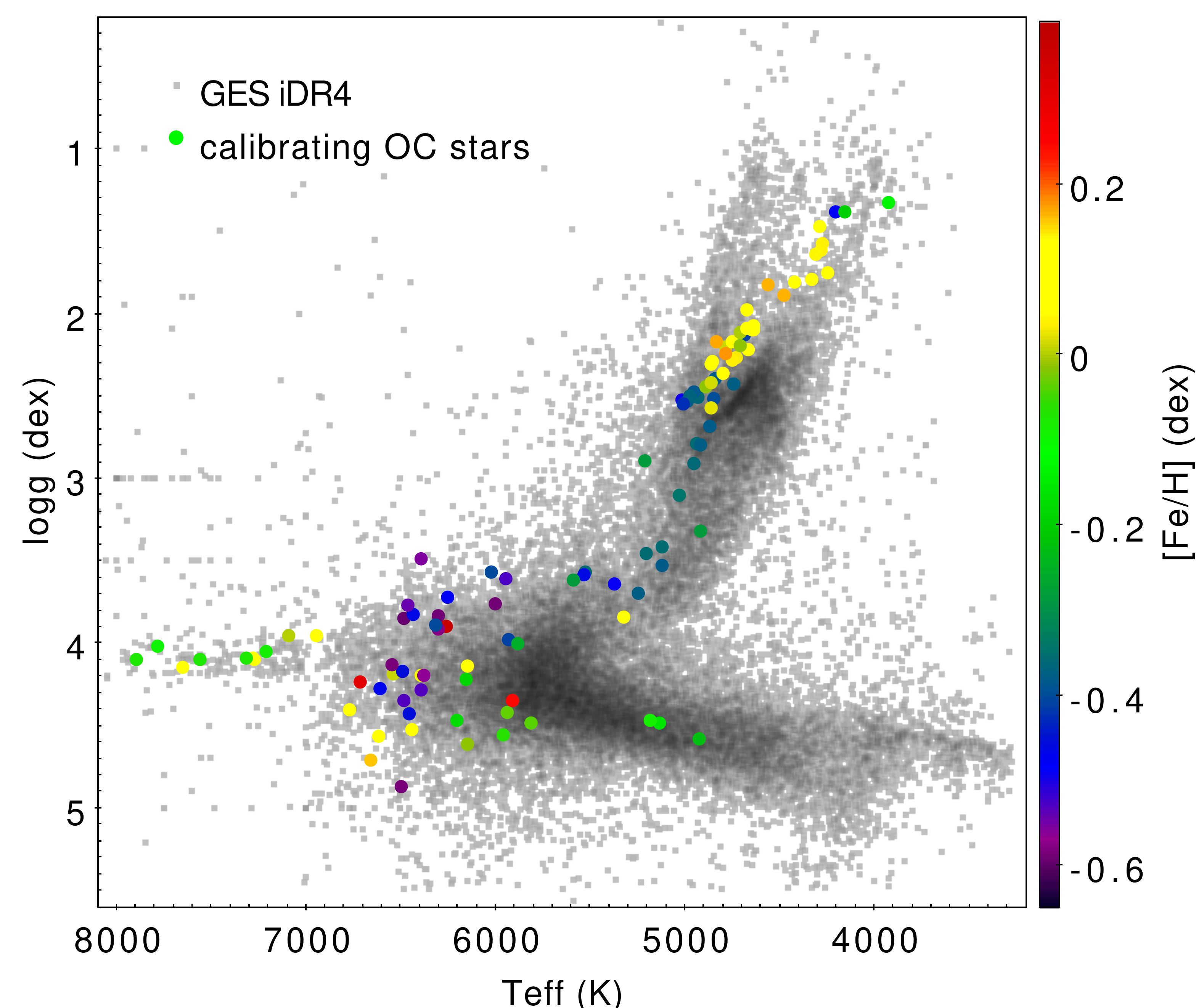} \caption{Position on the
      T$_{\rm{eff}}$-log$g$ plane of the stars in calibrating OCs including
      archival data analyzed in iDR4 (namely NGC~2243, NGC~6705, and NGC~3532) and
      coloured according to their [Fe/H]. These stars were observed with both the
      MW field and the OC setups, there are of course many more OC stars observed
      with the OC setups only (of the order of 20\,000, see Table~\ref{tab-setup}).
      The whole GES iDR4 sample is reported in the background as smaller grey
      dots.} \label{fig-ocstars}
   \end{figure}

\begin{table*}
\caption{GES calibrating globular clusters, with basic properties from the Harris
Galactic GC catalogue \citep{harris96,harris10}, except where noted. The status
column specifies the processing cycle in which each GC was analysed for the first
time (see Section~\ref{sec-strat}), and the last column indicates other surveys
using each GC as calibrator, along with other useful annotations.}            
\label{tab-gc}      
\centering          
\begin{tabular}{l c c c r r c l}     
\hline\hline       
Cluster          & [Fe/H] & E(B--V) & (m-M)$_V$ &$\langle$RV$\rangle$& $\sigma_0$ & Status    & Notes \\ 
                 & (dex)  & (mag)   & (mag)     &  (km/s)         & (km/s) & \\ 
\hline
 NGC 1851        & --1.18 &    0.02 &    15.47  &   320.5 &  10.4 & in iDR1   & GALAH \\
 NGC 4372        & --2.17 &    0.39 &    15.03  &    72.3 &   3.9\tablefootmark{a}    & in iDR1   & metal-poor \\
 NGC 5927        & --0.49 &    0.45 &    15.82  & --107.5 &   5.1\tablefootmark{a} & in iDR1   & metal-rich \\
 NGC 2808        & --1.14 &    0.22 &    15.59  &   101.6 &  13.4 & in iDR2   & well studied \\
 NGC 7078 (M 15) & --2.37 &    0.10 &    15.39  & --107.0 &  13.5 & in iDR2   & APOGEE, GALAH \\
 NGC 4833        & --1.85 &    0.32 &    15.08  &   200.2 &   3.9\tablefootmark{a}    & in iDR2   & metal-poor \\
 NGC 6752        & --1.54 &    0.02 &    13.13  &  --26.7 &   4.9 & in iDR3   & RAVE, GALAH \\
 NGC 104 (47 Tuc)& --0.72 &    0.04 &    13.37  &  --18.0 &  11.0 & in iDR3/4 & GALAH \\
 NGC 362         & --1.26 &    0.05 &    14.83  &   223.5 &   6.4 & in iDR4   & GALAH \\
 NGC 1904 (M 79) & --1.60 &    0.01 &    15.59  &   205.8 &   5.3 & in iDR4   & well studied \\
 NGC 7089 (M 2)  & --1.65 &    0.06 &    15.50  &   --5.3 &   8.2 & in iDR4   & APOGEE \\
 NGC 6553        & --0.18 &    0.63 &    15.83  &   --3.2 &   6.1 & observed  & metal-rich \\
 NGC 1261        & --1.27 &    0.01 &    16.09  &    68.2 &   ... & observed  & well studied \\
 NGC 6218 (M 12) & --1.37 &    0.19 &    14.01  &  --41.4 &   4.5 & observed  & RAVE \\
\hline \hline                
\multicolumn{8}{l}{$^{(a)}$Radial velocity dispersion from \citet{lardo15}.}\\
\end{tabular}                                         
\end{table*}

\subsection{Selection criteria for individual OC stars}

For OCs, the individual star selection criteria varied from case to case. The
reliable members observed with both the OC and field setups, and included in iDR4,
are displayed in Figure~\ref{fig-ocstars}. Our main guidelines were: 

\begin{itemize}
\item{to profit from the target selection effort performed by the GES OC group
(Bragaglia et al., in preparation), selecting the candidates among stars that
already had good membership information from the literature, or from previous
GES internal releases; in other words, for most calibrating OCs, the selected
stars are a subsample of those observed for scientific purposes;}
\item{to connect stars in different evolutionary phases, i.e., on the red clump
and on the MS, whenever it was possible to select MS stars in a convienient
magnitude range without including too many fast rotating stars in the sample;}
\item{to sample a range of APs to test the self-consistency of the abundance
analysis -- similarly to the case of GCs -- selecting stars in a range of 1--2
mag on the MS for those OCs for which a low fraction of fast rotators were
present in the affordable magnitude ranges; in those OCs, we selected stars
spanning a range of 1--2 magnitudes;}
\end{itemize}

Additionally, ESO FLAMES archival data of the relevant OCs will be included in
GES, as explained above: for example, many of the stars analysed in NGC~6705 or
M~67 come from the ESO UVES archive. It is important to note that scientific OC
observations can also be used by the WGs or by WG15 to homogenize the results. 


\subsection{GC selection criteria}
\label{sec-globular}

The selection of calibrating GCs\footnote{GCs are not part of the scientific
targets of GES, they are only observed for calibration purposes.} proceeded by
considering clusters that were used by other surveys like RAVE
\citep{rave1,rave2,rave3,lane11}, GALAH \citep[][Martell, 2014, private
communication]{galah}, and APOGEE
\citep{frinchaboy12,frinchaboy13,anders13,meszaros13}, or that were subject to
numerous high-resolution studies in the past. 

   \begin{figure}
   \centering
   \includegraphics[width=\hsize]{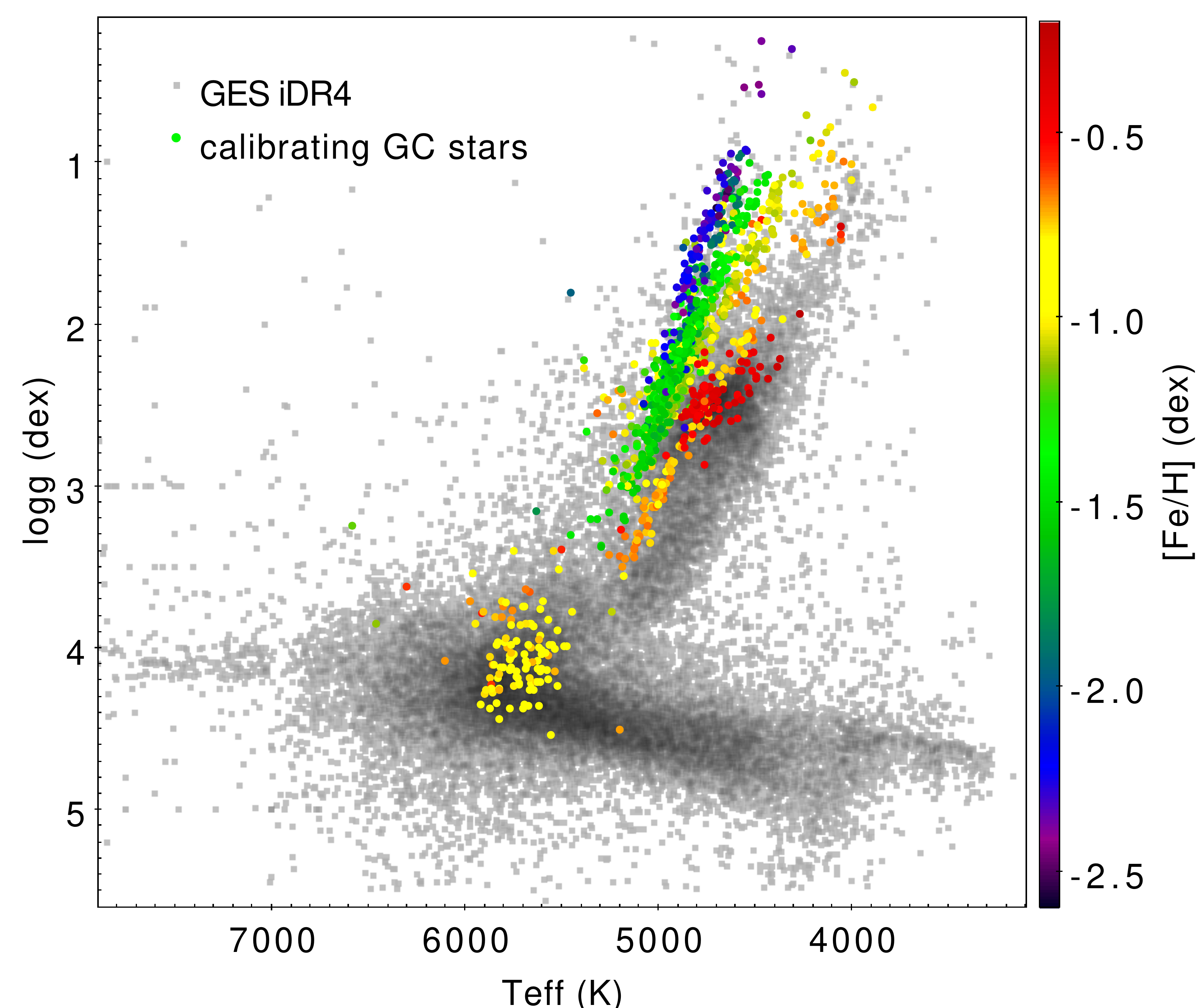} \caption{Position on the
      T$_{\rm{eff}}$-log$g$ plane of selected member stars in calibrating GCs,
      including archival data analyzed in iDR4 and coloured according to their
      [Fe/H]. The whole GES iDR4 sample is reported in the background as smaller
      grey squares.} \label{fig-gcstars}
   \end{figure}

Another fundamental criterion was the availability of wide field ($\simeq$
25$^{\prime}$, the FLAMES field of view), accurate photometric data in the
literature or in the archives. Unfortunately, at the time when GES started, not
many public photometric catalogues were available that covered the required
field of view. Therefore, we made use of the large amount of WFI (Wide Field
Imager) public GC data in the ESO archive. All relevant data were pre-reduced
with IRAF and then analyzed with DAOPHOT~II and ALLSTAR
\citep{daophot1,daophot2}, and the resulting magnitudes will be published in the
next public GES release. A more comprehensive set of photometric catalogues,
including data from all available public archives, is being prepared by
P.~Stetson\footnote{http://www3.cadc-ccda.hia-iha.nrc-cnrc.gc.ca/en/community/STET
SON/homogeneous/} and the catalogues for GES GCs will be published elsewhere. It
is important to stress that for dense stellar fields like GCs, the available
survey catalogues -- used to select GES targets for the MW field -- are
not precise enough. An example of the improvement that specific crowded-field
PSF-fitting techniques can bring over a standard photometric analysis was
presented by \citet{an08} concerning GCs in SDSS.

We thus created a sample including as many clusters as possible, selected from
the other surveys calibrating samples, that were visible from the South. We then
filled the gaps in [Fe/H] with clusters having public photometry data (from the
ESO archive or from the literature). Special care was taken in including
metal-rich GCs, as an interface with the OCs (see next Section) and considering
that the majority of GES field targets are relatively metal-rich. 

Twelve GCs were analyzed in iDR4, but two of them were not complete: NGC~4372
and NGC~6553. They will be fully included in subsequent releases, along with a
few more GCs. A complete list of observed GCs can be found in
Table~\ref{tab-gc}, with the metallicity coverage illustrated in
Figure~\ref{fig-gc}. 

\subsection{Selection criteria for individual GC stars}

We focused on red giants, because they are generally the best-studied objects in
the literature. A few subgiants and main sequence (MS) stars were previously
observed with UVES or GIRAFFE and were included in the GES analysis cycles along
with other relevant ESO archival data. For UVES, we avoided repeating stars that
already had good-quality UVES spectra in the archive. We did however repeat with
UVES a few of the stars having GIRAFFE observations in the archive, to build a
small sample of stars observed with both instruments for internal calibration
purposes. All the other UVES targets were high probability members, based on
their position in the CMD\footnote{During the first few GES runs, due to strict
scheduling requirements, we were forced to observe three clusters with high
differential reddening: NGC~4833, NGC~5927, and NGC~4372. For these, the
percentage of member stars among the selected targets was significantly lower
than for the other GCs. For NGC~5927 we could rely on a published study with RVs
of red clump stars \citep{simmerer13}, therefore in that case the majority of
selected stars turned out to be members. Anyway, even if field contaminants
cannot directly help for calibrations, they have an obvious scientific value for
GES. }. Stars with companions brighter than 1\% of their flux in a 1$'$ circle
were excluded from observations. 

For GIRAFFE, we gave highest priority to red giants having already archival
observations in non-GES HR setups (see Table~\ref{tab-setup}), because {\em (i)}
they were in most cases analyzed and published, so we had additional information
like RVs and chemistry to assess their membership and {\em (ii)} having a larger
wavelength coverage (with more GIRAFFE setups observed) can produce a more
reliable estimate of the APs and abundance ratios. Some effort was dedicated,
whenever possible, to observe a few stars in common with other spectroscopic
surveys mentioned above. Whenever additional membership information was
available in the literature (RVs, proper motions, metallicities), it was used to
select the most probable cluster members.

Depending on the GC and on the available body of archival data, the final sample
of stars analyzed in iDR4 per GC was of the order of 10--50 with UVES and
50--200 with GIRAFFE. All GES data were observed with UVES 580 and GIRAFFE HR10
and HR21. The candidates and archival data span a range of 1--3 magnitudes along
the red giant branch in each GC, which implies significant variations of APs in
stars with the same [Fe/H], thus allowing for rather precise tests on the
parameters and the self-consistency of the analysis (see below). To select
provisional members for this paper, we used a 3~$\sigma$ cut around the median
RV and [Fe/H] of the GES iDR4 recommended values, that were always fully
compatible with the literature reference values reported in Table~\ref{tab-gc}.
The position in the theoretical T$_{\rm{eff}}$-$\log g$ plane of the selected
members can be seen in Figure~\ref{fig-gcstars}.

   \begin{figure}
   \centering
   \includegraphics[width=\hsize]{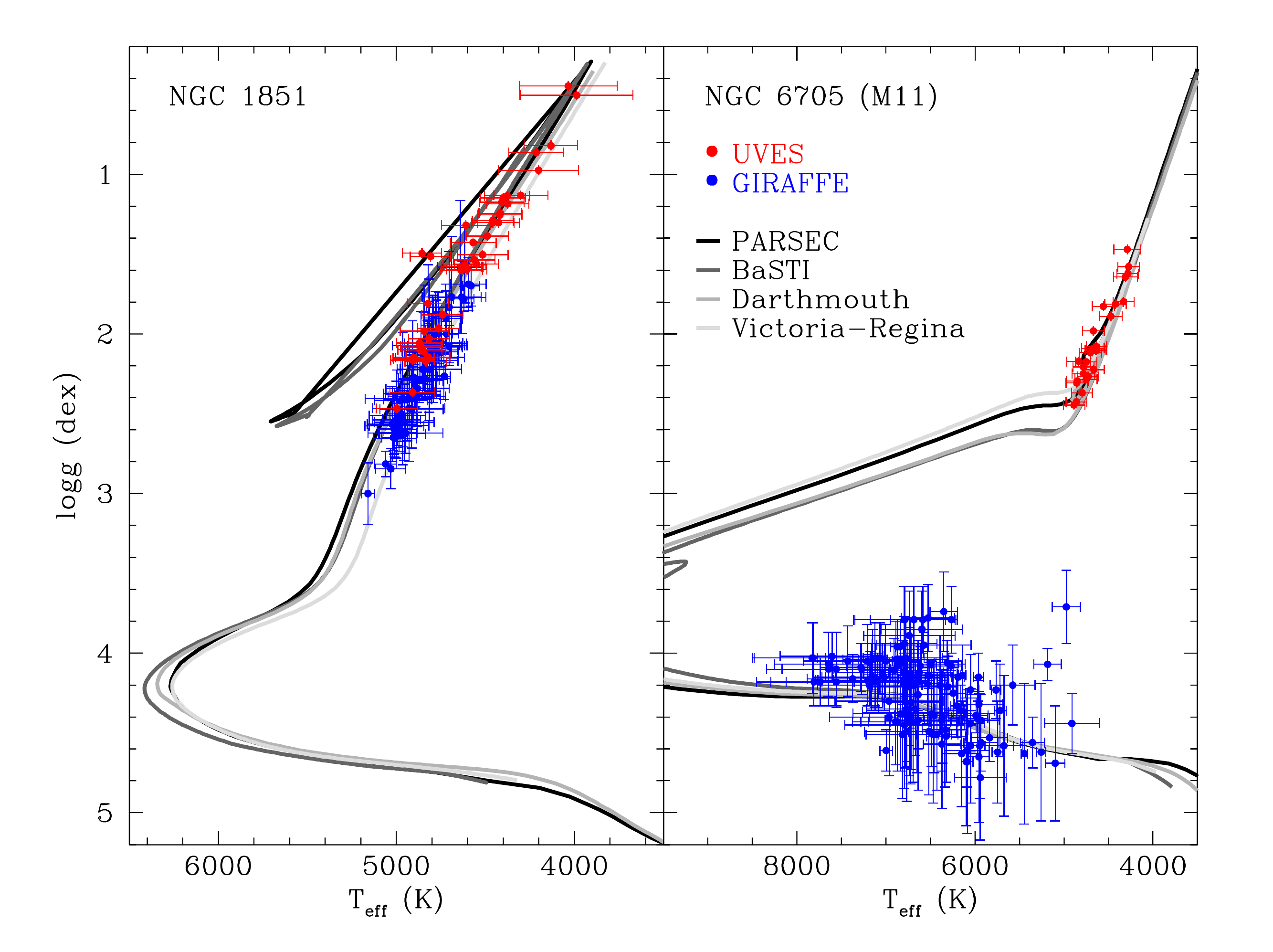}
      \caption{Example of comparison with theoretical models for NGC~1851 and
      NGC~6705 (M~11). All stars observed in NGC~6705 are plotted, even those
      that are observed only with the OC setups. UVES targets are plotted in red
      and GIRAFFE ones in blue. Four different sets of isochrones are plotted
      (see text for more details) as thick lines of different colours.} 
          \label{fig-iso}
   \end{figure}


\subsection{Selected results on calibrating clusters} 
\label{sec-clusterparams}

Clusters were used within GES past releases at many different levels, to
compare results obtained by different nodes, WGs, or observing setups, and to
study internal trends of abundances with APs. They were useful to identify
various problems that were later remedied in iDR4. Clusters were not, however,
used as {\em absolute} calibrators in iDR4, but rather were used {\em a
posteriori} to test the goodness of the overall homogenization process.
Therefore it is interesting to compare the final iDR4 cluster recommended values
with state-of-the-art external reference values, to give an idea of the results
of the whole GES homogeneization procedure.

   \begin{figure}
   \centering
   \includegraphics[width=\hsize]{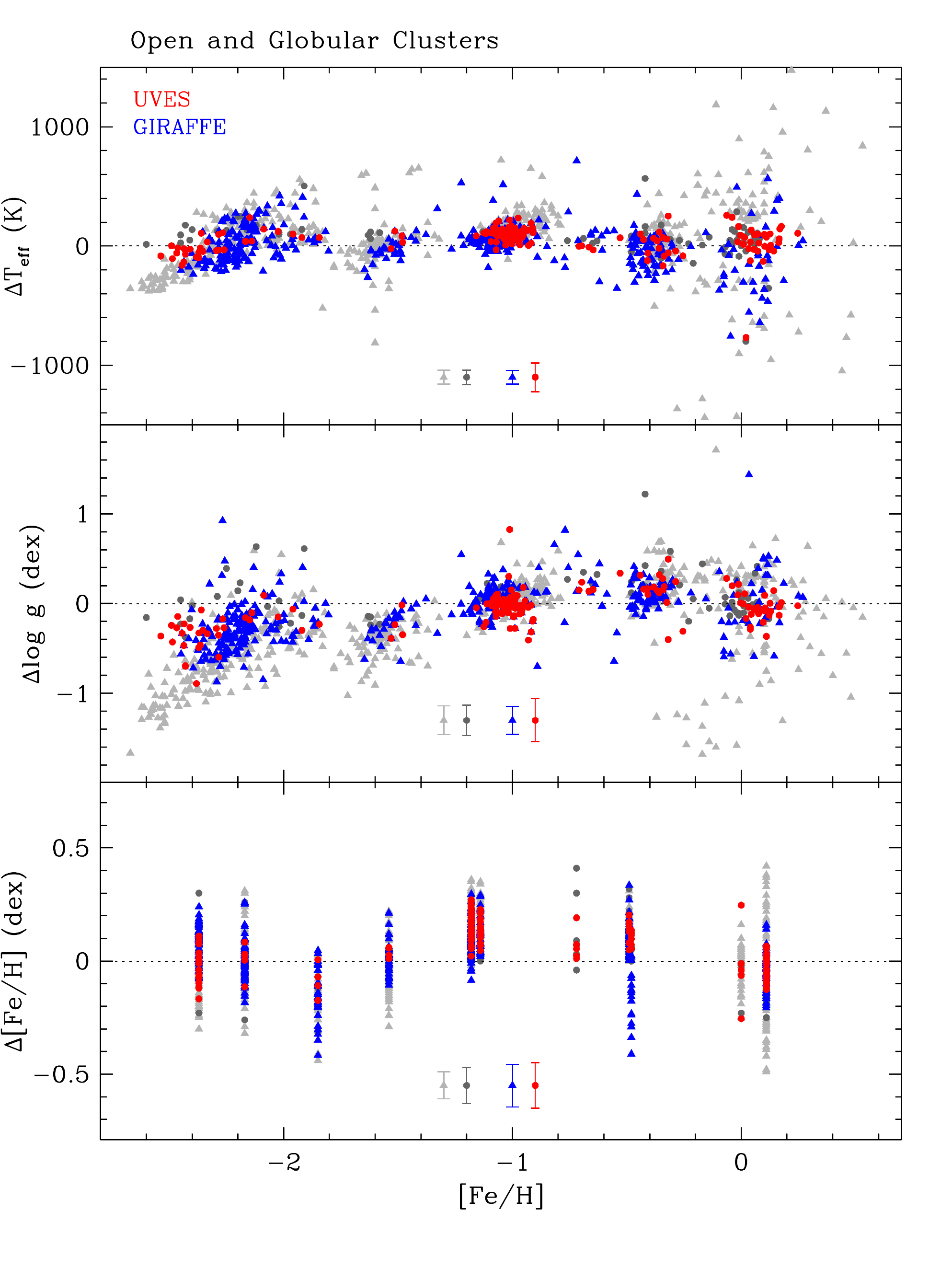}
      \caption{Difference between the GES recommended APs and [Fe/H] and the
      photometric reference APs and literature [Fe/H] (see text for more
      details). The GES iDR4 results are plotted as blue triangles (for GIRAFFE)
      and red circles (for UVES). Greyed out symbols represent the previous
      internal release (iDR2 + iDR3) corresponding values. The top panels shows
      $\Delta$T$_{\rm{eff}}$ as a function of GES [Fe/H], the middle panel
      $\Delta$log$g$ as a function of GES [Fe/H], and the bottom panel
      $\Delta$[Fe/H] as a function of the literature reference [Fe/H]. The
      median internal errors of GES recommended values are also plotted at the
      center of each panel.}
          \label{fig-phot}
   \end{figure}

A first comparison can be made with stellar models. In Figure~\ref{fig-iso} we
show NGC~1851 and NGC~6705 as an example. A more extensive discussion and set of
model comparisons will be presented in H17. We used four different sets of
stellar isochrones: the PARSEC set \citep{parsec1,parsec2,parsec3}; the BaSTI
set \citep{basti1,basti2}; the Dartmouth set \citep{dartmouth}; and the
Victoria-Regina set \citep{victoria}. We adopted the parameters listed in
Tables~\ref{tab-oc} and \ref{tab-gc}, with an age of 12.8~Gyr for NGC~1851.
As can be seen, apart from the small residual offset between the GIRAFFE and
UVES results (see below), the GES iDR4 recommended parameters agree well with
theoretical predictions, within the quoted uncertainties. Considering the
automated analysis, which is not tailored to obtain the best results for GCs,
this is a very satisfactory result.

For a different comparison, we computed independent T$_{\rm{eff}}$ and log$g$
values from our photometry, described in Section~\ref{sec-globular}, using the
\citet{alonso99} and \citet{alonso96} calibrations for giants and dwarfs,
respectively. To obtain T$_{\rm{eff}}$, we used the B--V and V--K colors,
dereddened with the E(B--V) values listed in Tables~\ref{tab-gc} and
\ref{tab-oc}, and we transformed the K$_{\rm{2MASS}}$ magnitudes into
K$_{\rm{TCS}}$ ones with the relations by \citet{ramirez05}. Similarly,
we obtained log$g$ using bolometric corrections from the cited calibrations and
fundamental relations. We assumed a fixed mass of 0.8~M$_{\odot}$ for evolved GC
stars and a varying mass for OC stars at various evolutionary stages, based on
the above selected isochrone sets.

The results of the comparison are presented in Figure~\ref{fig-phot}, where we
also show the results obtained during the previous internal processing cycle
(iDR2 + iDR3, based on data gathered in the first two years of GES
observations). As can be seen, there has been enormous improvement from the
previous to the present internal data release, especially at the two extremes of
the metallicity range. The causes of the improvement lie in the cyclic nature of
GES data analysis and calibration, where with each cycle not only new data are
added, but new procedures are introduced either to implement lessons learned in
previous cycles, or to refine the quality control and data analysis. A similar
analysis, based on the entire GES sample, will be presented in Randich et al.
(2016, in preparation)

The median iDR4 offsets to the reference values are always compatible with zero
-- within the uncertainties -- and the typical 1\,$\sigma$ spreads for both UVES
and GIRAFFE are compatible with the median GES errors. Of course, the selected
reference parameters depend on the chosen reference cluster parameters in
Tables~\ref{tab-gc} and \ref{tab-oc}, on the colour-temperature calibration
relations and their errors, on the accuracy and precision of the reference
photometry, and so on. Had we chosen the \citep{gonzalez09} colour-temperature
calibration, for example, the median T$_{\rm{eff}}$ differences reported below
would have been lower by about $\simeq$60~K. What is important to note here is
that the reference APs are derived with an independent method and yet the
agreement is quite satisfactory, especially considering that cluster stars in GES
are analyzed with the same method as field stars, i.e., without profiting from
the extra information on distance provided by clusters. For UVES we obtained
$<\Delta$T$_{\rm{eff}}>$ = 71 $\pm$ 93~K, $<\Delta \log g>$ = 0.04 $\pm$
0.18~dex, and $<\Delta$[Fe/H]$>$ = 0.06 $\pm$ 0.11~dex. For GIRAFFE we obtained
$<\Delta$T$_{\rm{eff}}>$ = --49 $\pm$ 149~K, $<\Delta \log g>$ = --0.21 $\pm$
0.30~dex, and $<\Delta$[Fe/H]$>$ = 0.00 $\pm$ 0.16~dex, where the quoted
uncertainties are 1\,$\sigma$ spreads. 


\section{Astroseismologic constraints}
\label{sec-corot}

The resonant frequencies of stochastically-driven pulsators (such as the Sun and
other FGK-type dwarfs and giants with turbulent convective envelopes) allow for
precise estimates of stellar APs that are largely independent of
spectroscopy \citep[see e.g.][and references therein]{miglio13}. As an example
the surface gravity $\log{g}$, a relatively difficult quantity to measure
directly from spectroscopy alone, is strongly correlated with  the frequency at
maximum oscillation power ($\nu_{\rm max}$): $$\nu_{\rm max} \propto
g/\sqrt(T_{\rm eff})$$ \noindent \citep{brown91,kjeldsen95,belkacem11}. Given the
typical accuracy of these scaling relations and the precision of the measured
$\nu_{\rm max}$, the seismic estimates of $\log g$ are likely more precise
($\sigma_{\log g}\sim 0.05$~dex) than those derived from standard spectroscopic
methods, that are typically in the range $\sigma_{\log g}\simeq 0.1-0.2$~dex.
Note also the weak dependence of $\log g$ on T$_{\rm eff}$: a shift in T$_{\rm
eff}$ of $\approx$ 100~K leads to an expected variation in $\log g$ of less than
0.01~dex, at least in the mass range covered by GES.

There is good agreement between the $\log g$ values inferred from seismology and
from classical methods for bright stars spanning a wide range of effective
temperature and evolutionary state \citep[dwarfs, sub-giants and red
giants,][]{morel12,morel14b}. This supports the application of scaling relations
in deriving weakly model-dependent $\log{g}$ estimates, at least for the tested
domains of metallicity and surface gravity. In the case of Kepler, the 
spectroscopic and seismic gravities have shown a good  agreement, with no
evidence of systematic offsets: $\langle \log g _{\rm {spec}} - \log g_{\rm
{seism}}\rangle = +0.08\pm0.07$~dex for dwarfs \citep{bruntt12} and
$-0.05\pm0.30$~dex for giants \citep{thygesen12}. Fixing $\log{g}$ to the seismic
value in spectroscopic analysis --- whenever possible --- has become an
increasingly popular technique \citep[e.g.,][]{huber13}. The availability of
precise seismic $\log{g}$ estimates for thousands of solar-like pulsators
detected by CoRoT \citep{michel08} and Kepler \citep{borucki10} missions makes
them valuable targets for science verification and/or calibration. 

\begin{figure}
\includegraphics[width=\columnwidth]{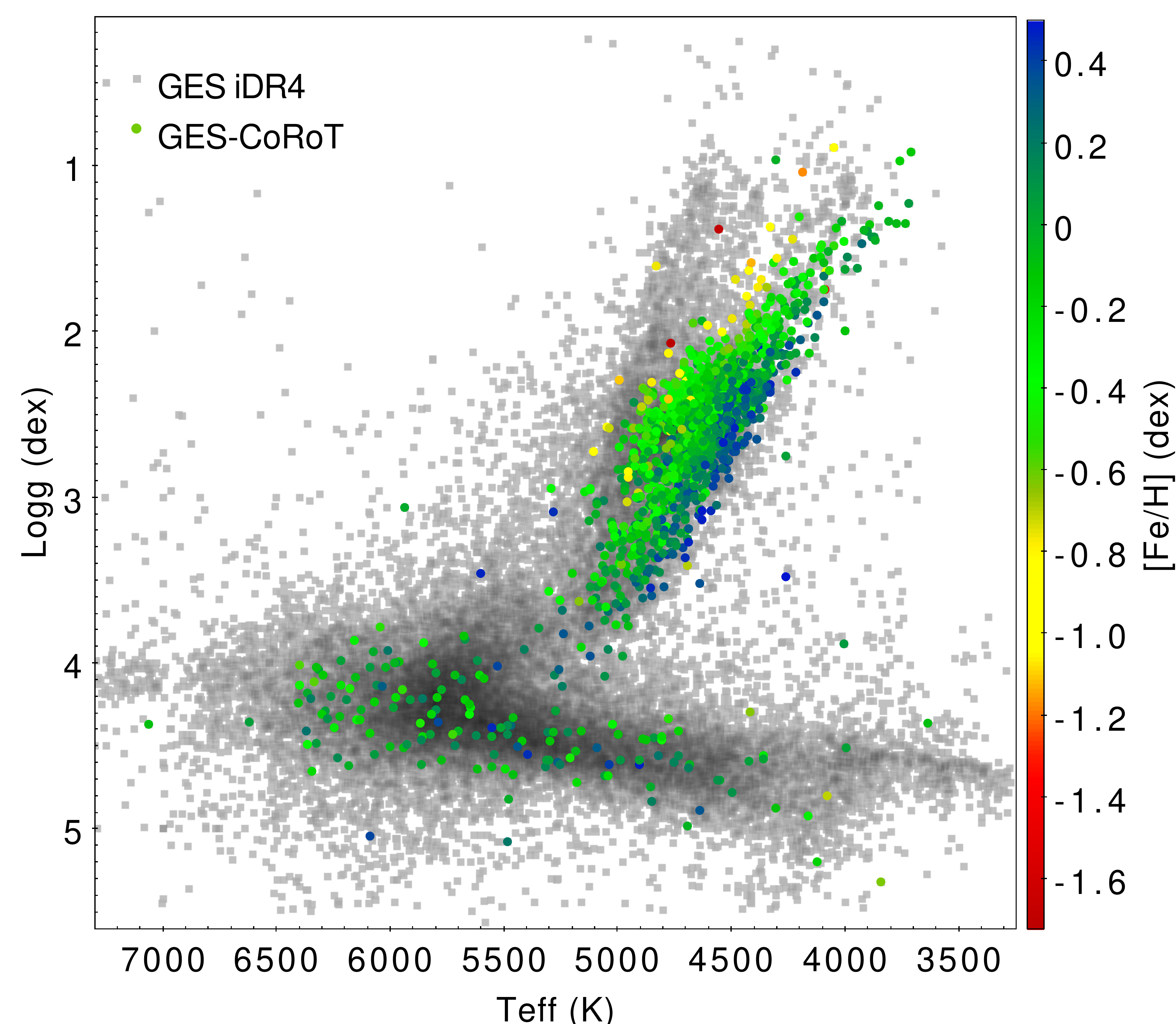}
\caption{Position in the T$_{\rm{eff}}$-log$g$ plane of the GES targets -- in the
direction of the CoRoT center and anti-center fields -- that were analyzed in
iDR4, coloured according to their [Fe/H]. The GES iDR4 analysis is not the final
result, because the GES-CoRoT project is still ongoing (see text for more
details). The whole GES iDR4 sample is reported in the background as smaller grey
squares.} 
\label{fig-corot}
\end{figure}

\subsection{Asteroseismic collaborations with GES}
\label{sec-ges-astereo}

GES observed selected targets in the LRc01 and LRa01 CoRoT fields (in the
Galactic center and anticenter directions, respectively) where CoRoT has
detected and characterized more than 2\,000 oscillating G-K red giants
\citep[][see also Figure~\ref{fig-corot}]{mosser10}. More than 1\,500 red giants
were observed with the GES field setups (Table~\ref{tab-setup}) and analysed in
iDR4. A subset of a few tens of the candidates, for which the oscillation
spectra have also allowed to derive their evolutionary state \citep[either RGB
or central He-burning,][]{mosser11}, were observed with UVES. The GES-CoRoT
collaboration will provide a set of reference parameters to compare with the GES
recommended parameters, similarly to what was done by \citet{jofre14} for the
benchmark stars. The final reference APs for these stars will be derived by a
combined team of GES and CoRoT scientists, after an iterative process: the
spectroscopic T$_{\rm eff}$ value obtained by GES and the seismic parameter
$\nu_{\rm max}$ will provide a first seismic $\log g$ value, which will be held
fixed in the following spectroscopic analysis by a subset of the GES abundance
analysis nodes participating to the GES-CoRoT project. The new T$_{\rm eff}$
value will then provide a new seismic $\log g$ estimate, and so on, until
convergence (typically no more than two iterations are needed). The results of
this project will be presented elsewhere.

\begin{figure}
\includegraphics[width=\columnwidth]{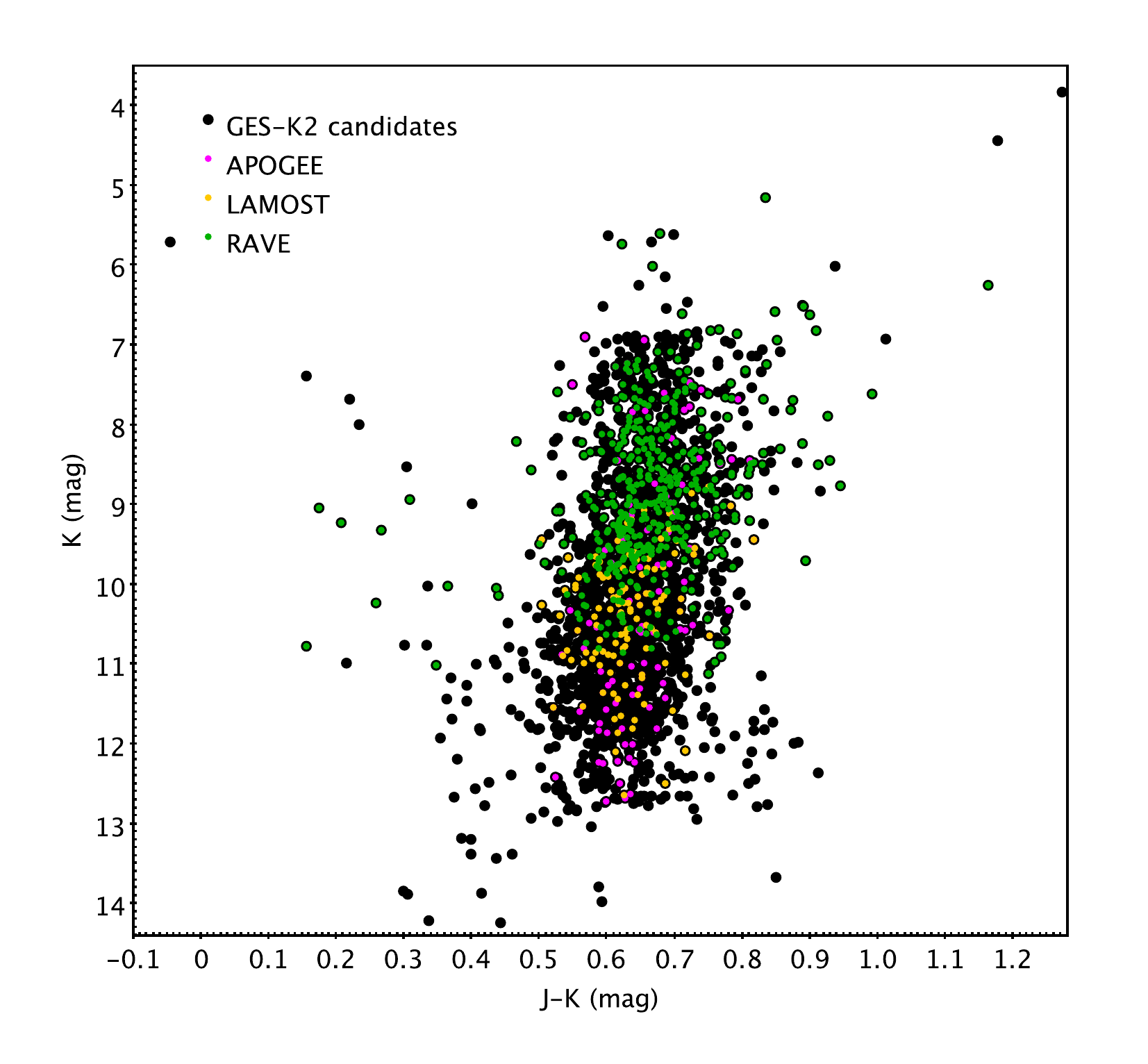}
\caption{Kepler red giants in the K2 C1 and C3 fields, that were selected as
candidates for GES observations (large black circles). Stars already observed
by other surveys are highlighted in different colors (APOGEE in
magenta, LAMOST in yellow, and RAVE in green).} 
\label{fig-k2}
\end{figure}

Similarly, thousands of solar-like oscillating giant stars have been discovered
in Campaigns 1 and 3 of the K2 mission. GES observations are currently planned
for a combined asteroseismic-spectroscopic analysis using individual resonant
frequencies with the Birmingham asteroseismic group, allowing for far more
insight on the physics of stellar interiors than what is available using simple
scaling relations \citep[e.g.,][]{davies16}. Other spectroscopic surveys are
targeting giants observed by Kepler and CoRoT for similar purposes, so a large
sample of overlapping spectroscopic observations is expected (see also
Figure~\ref{fig-k2}), allowing for future survey intercalibrations. More details
on the target selection strategy, data analysis, and use of these calibrators
for the intercalibration with other surveys can be found in Gilmore et al.,
(2016, in preparation).

\section{Discussion and conclusions}
\label{sec-concl}

The GES calibration and data analysis strategy is designed to take care of the
internal consistency and the overall robustness of its results with respect to
literature or reference values. The abundance analysis process in GES is
complex, resulting from observations of different objects with different
instrumental setups, and analyzed by several abundance analysis nodes, using
virtually all of the existing state-of-the-art techniques. While this is one of
the major strengths of the GES data analysis, it requires particular attention
in the process of data homogenisation, that produces the GES recommended RVs,
APs, and chemical abundance ratios. 

Different classes of calibrating objects were selected, with the main goals of
covering the different observational setups, the AP space covered by the GES
scientific targets, and the variety of methods used to analyze them. In
particular, we selected a sample of 31 RV standards from the Gaia RV standards
catalogue \citep{soubiran13}; a pool of star clusters, both GCs and OCs, either
used as calibrators by other major ongoing surveys or well studied in the
literature, of which 21 were observed to date; a list of FGK benchmark stars in
common with the Gaia list of benchmarks \citep{heiter15b} was observed,
complemented by cooler M benchmarks and OBA candidate benchmark stars; and a
list of thousands of targets in common with those of the two major astroseismic
space missions, CoRoT and Kepler, was also observed. In a few cases the
calibration planning of GES and its requirements have spawned calibration
projects like the Gaia benchmarks spectroscopic project
\citep{blanco14,jofre14,jofre15,hawkins16} or the GES-CoRoT collaboration, that
will prove useful for many other projects and surveys. 

The complex GES calibration and homogenization procedures, described in
\citet{smiljanic14}, \citet{lanzafame15}, and H17, are applied at different
levels of the data processing (node, WG, and survey-wide) and they are applied to
all the GES targets (field stars, OC scientific targets, and calibrators).
Therefore, it is particularly instructive to examine the outcome of the whole
calibration process on the calibrating objects themselves. We presented a few
examples of the comparisons that are routinely performed in GES. In particular,
we showed how the cyclic processing leads to significant improvement from cycle
to cycle. We also quantitatively showed that the agreement between GES iDR4
recommended values and reference values for the calibrating objects are very
satisfactory. The average offsets and spreads are generally compatible with the
GES measurement errors, proving that the performance goals set by
\citet{gilmore12} and \citet{randich12} are being met.


\begin{acknowledgements}

Based on data products from observations made with ESO Telescopes at the La Silla
Paranal Observatory under programme ID 188.B-3002 and 193.B-0936. These data
products have been processed by the Cambridge Astronomy Survey Unit (CASU) at the
Institute of Astronomy, University of Cambridge, and by the FLAMES/UVES reduction
team at INAF--Osservatorio Astrofisico di Arcetri. These data have been obtained
from the Gaia-ESO Survey Data Archive, prepared and hosted by the Wide Field
Astronomy Unit, Institute for Astronomy, University of Edinburgh, which is funded
by the UK Science and Technology Facilities Council. This work was partly
supported by the European Union FP7 programme through ERC grant number 320360 and
by the Leverhulme Trust through grant RPG-2012-541. We acknowledge the support
from INAF and Ministero dell'Istruzione, dell'Universit\`a e della Ricerca (MIUR)
in the form of the grant "Premiale VLT 2012". The results presented here benefit
from discussions held during the Gaia-ESO workshops and conferences supported by
the ESF (European Science Foundation) through the GREAT Research Network
Programme. S.F. and T.B. acknowledge the support from the New Milky Way project
funded by a grant from the Knut and Alice Wallenberg foundation. C.L. gratefully
acknowledges financial support from the European Research Council
(ERC-CoG-646928, Multi-Pop, PI: N. Bastian). U.H. and A.J.K acknowledge support
from the Swedish National Space Board (Rymdstyrelsen). The research of A.L. been
subsidized by the Belgian Federal Science Policy Office under contract No.
BR/143/A2/BRASS. R.S. acknowledges support by the National Science Center of
Poland through grant 2014/15/B/ST9/03981. C.A.P. is thankful for support from the
Spanish Ministry of Economy and Competitiveness (MINECO) through grant
AYA2014-56359-P. J.M. acknowledges support from the ERC Consolidator Grant
funding scheme (project STARKEY, G.A. n.615604). T.M. acknowledges financial
support from Belspo for contract PRODEX GAIA-DPAC. S.G.S acknowledges the support
by Funda\c c\~ao para a Ci\^encia e Tecnologia (FCT) through national funds and a
research grant (project ref. UID/FIS/04434/2013, and PTDC/FIS-AST/7073/2014).
S.G.S. also acknowledge the support from FCT through Investigador FCT contract of
reference IF/00028/2014 and POPH/FSE (EC) by FEDER funding through the program
``Programa Operacional de Factores de Competitividade -- COMPETE". L.S.
acknowledges support by the Ministry of Economy, Development, and Tourism's
Millennium Science Initiative through grant IC120009, awarded to The Millennium
Institute of Astrophysics (MAS). M.Z. acknowledges support  by the Ministry of
Economy, Development, and Tourism's Millennium Science Initiative through grant
IC120009, awarded to The Millennium Institute of Astrophysics (MAS), by Fondecyt
Regular 1150345 and by the BASAL CATA PFB-06. E.J.A. and M.T.C acknowledge the
financial support from the Spanish Ministerio de Econom\'\i a y Competitividad,
through grant AYA2013-40611-P. S.Z. acknowledge the support from the INAF grant
"PRIN INAF 2014", "Star won't tell their ages to Gaia, Galactic Archaelogy with
wide-area asterosismic". This research has made use of the WEBDA database,
operated at the Department of Theoretical Physics and Astrophysics of the Masaryk
University; of the TOPCAT catalogue handling and plotting tool \citep{topcat}; of
the Simbad database and the VizieR catalogue access tool, CDS, Strasbourg, France
\citep{vizier}; and of NASA's Astrophysics Data System.

\end{acknowledgements}



\begin{thebibliography}{99}

\bibitem[Adelman et al.(1980)]{adelman80} Adelman, S.~J., White, R.~E., \&
Pyper, D.~M.\ 1980, \apjs, 43, 491 

\bibitem[Alonso et al.(1996)]{alonso96} Alonso, A., Arribas, S., \& Martinez-Roger,
C.\ 1996, \aap, 313, 873 

\bibitem[Alonso et al.(1999)]{alonso99} Alonso, A., Arribas, S., \&
Mart{\'{\i}}nez-Roger, C.\ 1999, \aaps, 140, 261 

\bibitem[An et al.(2008)]{an08} An, D., Johnson, J.~A.,  Clem, J.~L., et al.
\ 2008, \apjs, 179, 326 

\bibitem[Anders et al.(2013)]{anders13} Anders, F., Chiappini,  C., Santiago,
B.~X., et al.\ 2013, arXiv:1311.4549 



\bibitem[Anthony-Twarog \& Twarog(2000)]{twarog00} Anthony-Twarog, B.~J., 
\& Twarog, B.~A.\ 2000, \aj, 119, 2282 

\bibitem[Bailer-Jones et al.(2013)]{bailerjones13} Bailer-Jones, C.~A.~L.,
Andrae, R., Arcay, B., et al.\ 2013, \aap, 559, A74 

\bibitem[Balcells et al.(2010)]{balcells10} Balcells, M., Benn, C.~R., Carter,
D., et al.\ 2010, \procspie, 7735, 77357G 

\bibitem[Ballester et al.(2000)]{uves} Ballester, P., Modigliani, A., Boitquin,
O., et al.\ 2000, The Messenger, 101, 31 

\bibitem[Barb{\'a} et al.(2010)]{barba10} Barb{\'a}, R.~H., Gamen, R., Arias,
J.~I., et al.\ 2010, Revista Mexicana de Astronomia y Astrofisica Conference
Series, 38, 30 

\bibitem[Barb{\'a} et al.(2014)]{barba14} Barb{\'a}, R., Gamen, R., Arias,
J.~I., et al.\ 2014, Revista Mexicana de Astronomia y Astrofisica Conference
Series, 44, 148 

\bibitem[Barden et al.(2010)]{barden10} Barden, S.~C., Jones, D.~J., Barnes,
S.~I., et al.\ 2010, \procspie, 7735, 773509 

\bibitem[Barrado y Navascu{\'e}s et al.(2001)]{barrado01} Barrado 
y Navascu{\'e}s, D., Deliyannis, C.~P., \& Stauffer, J.~R.\ 2001, \apj, 549, 452 

\bibitem[Belkacem et al.(2011)]{belkacem11} Belkacem, K., Goupil, M.~J., Dupret,
M.~A., et al.\ 2011, \aap, 530, AA142 

\bibitem[Bensby et al.(2014)]{bensby14} Bensby, T., Feltzing, S., \& Oey, M.~S.\
2014, \aap, 562, AA71 

\bibitem[Bertaux et  al.(2014)]{tapas} Bertaux, J.~L., Lallement, R., Ferron, S.,
Boonne, C., \& Bodichon, R.\ 2014, \aap, 564, AA46 


\bibitem[Bikmaev et al.(2002)]{bikmaev02} Bikmaev, I.~F., Ryabchikova, T.~A.,
Bruntt, H., et al.\ 2002, \aap, 389, 537 

\bibitem[Blanco-Cuaresma et al.(2014)]{blanco14} Blanco-Cuaresma, S., Soubiran,
C., Jofr{\'e}, P., \& Heiter, U.\ 2014, \aap, 566, AA98 

\bibitem[Bland-Hawthorn et al.(2010)]{bland10} Bland-Hawthorn, J., Krumholz,
M.~R., \& Freeman, K.\ 2010, \apj, 713, 166 

\bibitem[Boyajian et al.(2012)]{boyajian12} Boyajian, T.~S., von Braun, K., van
Belle, G., et al.\ 2012, \apj, 757, 112 

\bibitem[Borucki et al.(2010)]{borucki10} Borucki, W.~J., Koch,  D., Basri, G.,
et al.\ 2010, Science, 327, 977 


\bibitem[Bragaglia \& Tosi(2006)]{bocce1} Bragaglia, A., \& Tosi, M.\ 2006, \aj, 
131, 1544 

\bibitem[Bressan et al.(2012)]{parsec1} Bressan, A., Marigo, P., Girardi, L., et
al.\ 2012, \mnras, 427, 127 


\bibitem[Brown(1991)]{brown91} Brown, T.~M.\ 1991, \apj, 371, 396 

\bibitem[Bruntt et al.(2012)]{bruntt12} Bruntt, H., Basu, S.,  Smalley, B., et
al.\ 2012, \mnras, 423, 122 

\bibitem[Burkhart \& Coupry(1989)]{burkhart89} Burkhart, C., \& Coupry, M.~F.\
1989, \aap, 220, 197 

\bibitem[Carraro et al.(2002)]{carraro02} Carraro, G., Girardi, L., \& Marigo, 
P.\ 2002, \mnras, 332, 705 

\bibitem[Chen et al.(2014)]{parsec2} Chen, Y., Girardi, L., Bressan, A., et al.\
2014, \mnras, 444, 2525 

\bibitem[Chubak et al.(2012)]{chubak12} Chubak, C., Marcy, G., Fischer, D.~A., et
al.\ 2012, arXiv:1207.6212 

\bibitem[Crowther et  al.(2006)]{crowther06} Crowther, P.~A., Lennon, D.~J., \&
Walborn, N.~R.\ 2006, \aap, 446, 279 

\bibitem[Daflon \& Cunha(2004)]{daflon04} Daflon, S., \& Cunha, K.\ 2004, \apj,
617, 1115 

\bibitem[Davies et al.(2016)]{davies16} Davies, G.~R., Silva Aguirre, V.,
Bedding, T.~R., et al.\ 2016, \mnras, 456, 2183 

\bibitem[de Bruijne(2012)]{gaia3} de Bruijne, J.~H.~J.\ 2012, \apss, 341, 31 

\bibitem[de Jong et al.(2014)]{dejong14} de Jong, R.~S., Barden,  S.,
Bellido-Tirado, O., et al.\ 2014, \procspie, 9147, 91470M 

\bibitem[de Laverny et al.(2012)]{delaverny12} de Laverny, P., Recio-Blanco, A.,
Worley, C.~C., \& Plez, B.\ 2012, \aap, 544, AA126 

\bibitem[De Silva et al.(2015)]{galah} De Silva, G.~M., Freeman, K.~C., Bland-Hawthorn, J., et al.\ 2015, \mnras, 449, 2604 

\bibitem[Demory et al.(2009)]{demory09} Demory, B.-O., S{\'e}gransan, D.,
Forveille, T., et al.\ 2009, \aap, 505, 205 

\bibitem[De Pascale et al.(2014)]{depascale14} De Pascale, M., Worley, C.~C., de
Laverny, P., et al.\ 2014, \aap, 570, AA68 

\bibitem[Dias et al.(2002)]{dias02} Dias, W.~S., Alessi, B.~S., Moitinho, A., \&
L{\'e}pine, J.~R.~D.\ 2002, \aap, 389, 871 

\bibitem[Donati et al.(2014)]{donati14} Donati, P., Cantat Gaudin, T.,
Bragaglia, A., et al.\ 2014, \aap, 561, AA94 

\bibitem[Dotter et al.(2008)]{dartmouth} Dotter, A., Chaboyer, B.,
Jevremovi{\'c}, D., et al.\ 2008, \apjs, 178, 89 

\bibitem[Ekstr{\"o}m et al.(2012)]{ekstrom12} Ekstr{\"o}m, S., Georgy, C.,
Eggenberger, P., et al.\ 2012, \aap, 537, AA146 

\bibitem[Freeman \& Bland-Hawthorn(2002)]{freeman02} Freeman, K., \&
Bland-Hawthorn, J.\ 2002, \araa, 40, 487 

\bibitem[Frinchaboy et al.(2013)]{frinchaboy13} Frinchaboy, P.~M., Thompson, B.,
Jackson, K.~M., et al.\ 2013, \apjl, 777, L1 

\bibitem[Frinchaboy et al.(2012)]{frinchaboy12} Frinchaboy, P.~M., Allende
Prieto, C., Beers, T.~C., et al.\ 2012, American Astronomical Society Meeting
Abstracts \#219, 219, \#428.04 

\bibitem[Gaia Collaboration (2016a)]{gaia1} Gaia Collaboration 2016,
arXiv:1609.04153 

\bibitem[Gaia Collaboration et al.(2016b)]{gaia2} Gaia Collaboration, Brown,
A.~G.~A., Vallenari, A., et al.\ 2016, arXiv:1609.04172 


\bibitem[Gilmore et al.(2012)]{gilmore12} Gilmore, G., Randich, S., Asplund, M.,
et al.\ 2012, The Messenger, 147, 25 

\bibitem[Gonz{\'a}lez Hern{\'a}ndez \& Bonifacio(2009)]{gonzalez09} Gonz{\'a}lez
Hern{\'a}ndez, J.~I., \& Bonifacio, P.\ 2009, \aap, 497, 497 

\bibitem[Gratton et al.(2012)]{gratton12} Gratton, R.~G., Carretta, E., \& Bragaglia,
A.\ 2012, \aapr, 20, 50 

\bibitem[Gray et al.(2001)]{gray01} Gray, R.~O., Graham, P.~W., \& Hoyt, S.~R.\
2001, \aj, 121, 2159 

\bibitem[Guarcello et al.(2007)]{guarcello07} Guarcello, M.~G., Prisinzano, L.,
Micela, G., et al.\ 2007, \aap, 462, 245 

\bibitem[Guarcello et al.(2010)]{guarcello10} Guarcello, M.~G., Micela, G.,
Peres, G., Prisinzano, L., \& Sciortino, S.\ 2010, \aap, 521, AA61 


\bibitem[Gustafsson et al.(2008)]{marcs} Gustafsson, B., Edvardsson, B.,
Eriksson, K., et al.\ 2008, \aap, 486, 951 

\bibitem[Harris(1996)]{harris96} Harris, W.~E.\ 1996, \aj, 112,  1487 

\bibitem[Harris(2010)]{harris10} Harris, W.~E.\ 2010, arXiv:1012.3224 

\bibitem[Heiter et al.(2014)]{heiter14} Heiter, U., Soubiran, C., Netopil, M., \&
Paunzen, E.\ 2014, \aap, 561, A93 

\bibitem[Heiter et al.(2015a)]{heiter15a} Heiter, U., Lind, K., Asplund, M., et
al.\ 2015a, \physscr, 90, 054010 

\bibitem[Heiter et al.(2015b)]{heiter15b} Heiter, U., Jofr{\'e}, P., Gustafsson,
B., et al.\ 2015b, \aap, 582, A49 

\bibitem[Hill(1995)]{hill95} Hill, G.~M.\ 1995, \aap, 294, 536 

\bibitem[Hawkins et al.(2016)]{hawkins16} Hawkins, K., Jofre, P., Heiter, U., et
al.\ 2016, arXiv:1605.08229 


\bibitem[Huber et al.(2013)]{huber13} Huber, D., Chaplin, W.~J.,
Christensen-Dalsgaard, J., et al.\ 2013, \apj, 767, 127 

\bibitem[Hubrig et al.(2008)]{hubrig08} Hubrig, S., Briquet, M., Morel, T., et
al.\ 2008, \aap, 488, 287 

\bibitem[Irrgang et al.(2014)]{irrgang14} Irrgang, A., Przybilla, N., Heber, U.,
et al.\ 2014, \aap, 565, AA63 

\bibitem[Jackson et al.(2015)]{jackson15} Jackson, R.~J., Jeffries, R.~D., Lewis,
J., et al.\ 2015, \aap, 580, A75 

\bibitem[Jeffries et al.(2006)]{jeffries06} Jeffries, R.~D.,  Maxted, P.~F.~L.,
Oliveira, J.~M., \& Naylor, T.\ 2006, \mnras, 371, L6 

\bibitem[Jeffries et al.(2014)]{jeffries14} Jeffries, R.~D., Jackson, R.~J.,
Cottaar, M., et al.\ 2014, \aap, 563, AA94 

\bibitem[Jofr{\'e} et al.(2014)]{jofre14} Jofr{\'e}, P., Heiter, U., Soubiran,
C., et al.\ 2014, \aap, 564, AA133 

\bibitem[Jofr{\'e} et al.(2015)]{jofre15} Jofr{\'e}, P., Heiter, U., Soubiran,
C., et al.\ 2015, \aap, 582, A81 

\bibitem[Kassis et al.(1997)]{kassis97} Kassis, M., Janes, K.~A., Friel, E.~D., 
\& Phelps, R.~L.\ 1997, \aj, 113, 1723 

\bibitem[Kjeldsen \& Bedding(1995)]{kjeldsen95} Kjeldsen, H., \& Bedding, T.~R.\
1995, \aap, 293, 87 

\bibitem[Koposov et al.(2011)]{koposov11} Koposov, S.~E.,  Gilmore, G., Walker,
M.~G., et al.\ 2011, \apj, 736, 146 

\bibitem[Kordopatis et al.(2013)]{kordopatis13} Kordopatis, G., Gilmore, G.,
Steinmetz, M., et al.\ 2013, \aj, 146, 134 

\bibitem[Lane et al.(2011)]{lane11} Lane, R.~R., Kiss, L.~L., Lewis, G.~F., et
al.\ 2011, \aap, 530, A31 

\bibitem[Lanzafame et al.(2015)]{lanzafame15} Lanzafame, A.~C., Frasca, A.,
Damiani, F., et al.\ 2015, \aap, 576, A80 

\bibitem[Lardo et al.(2015)]{lardo15} Lardo, C., Pancino, E., Bellazzini, M.,
et al.\ 2015, \aap, 573, AA115 

\bibitem[Lebzelter et al.(2012)]{lebzelter12} Lebzelter, T., Seifahrt, A.,
Uttenthaler, S., et al.\ 2012, \aap, 539, AA109 

\bibitem[Lefever et al.(2010)]{lefever10} Lefever, K., Puls, J., Morel, T., et
al.\ 2010, \aap, 515, AA74 


\bibitem[Lindegren \& Perryman(1996)]{lindegren96} Lindegren, L., \& Perryman,
M.~A.~C.\ 1996, \aaps, 116, 579 

\bibitem[Lindgren et al.(2016)]{lindgren16} Lindgren, S., Heiter, U., \&
Seifahrt, A.\ 2016, \aap, 586, A100 

\bibitem[Luck(1994)]{luck94} Luck, R.~E.\ 1994, \apjs, 91, 309 

\bibitem[Ma{\'{\i}}z Apell{\'a}niz et al.(2015)]{maiz15} Ma{\'{\i}}z
Apell{\'a}niz, J., Alfaro, E.~J., Arias, J.~I., et al.\ 2015, Highlights of
Spanish Astrophysics VIII, 603 


\bibitem[Majewski et al.(2015)]{majewski15} Majewski, S.~R., Schiavon, R.~P.,
Frinchaboy, P.~M., et al.\ 2015, arXiv:1509.05420 

\bibitem[Markova et al.(2014)]{markova14} Markova, N., Puls, J.,
Sim{\'o}n-D{\'{\i}}az, S., et al.\ 2014, \aap, 562, A37 

\bibitem[Martins et al.(2012)]{martins12} Martins, F., Escolano, C., Wade,
G.~A., et al.\ 2012, \aap, 538, AA29 

\bibitem[Martins et al.(2015)]{martins15} Martins, F., Herv{\'e}, A., Bouret,
J.-C., et al.\ 2015, \aap, 575, A34 

\bibitem[M{\'e}sz{\'a}ros et al.(2013)]{meszaros13}  M{\'e}sz{\'a}ros, S.,
Holtzman, J., Garc{\'{\i}}a P{\'e}rez, A.~E., et al.\  2013, \aj, 146, 133 

\bibitem[Michel et al.(2008)]{michel08} Michel, E., Baglin, A.,  Auvergne, M.,
et al.\ 2008, Science, 322, 558 

\bibitem[Miglio et al.(2013)]{miglio13} Miglio, A., Chiappini,  C., Morel, T.,
et al.\ 2013, European Physical Journal Web of Conferences,  43, 03004 


\bibitem[Mignard(2005)]{mignard05} Mignard, F.\ 2005, Astrometry in the Age of
the Next Generation of Large Telescopes, 338, 15 

\bibitem[Modigliani et al.(2004)]{flames2} Modigliani, A., Mulas, G., Porceddu,
I., et al.\ 2004, The Messenger, 118, 8 

\bibitem[Mokiem et al.(2005)]{mokiem05} Mokiem, M.~R., de Koter, A., Puls, J., et
al.\ 2005, \aap, 441, 711 


\bibitem[Morel \& Butler(2008)]{morel08a} Morel, T., \& Butler, K.\ 2008, \aap,
487, 307 


\bibitem[Morel \& Miglio(2012)]{morel12} Morel, T., \& Miglio, A.\ 2012, \mnras,
419, L34 


\bibitem[Morel et al.(2014)]{morel14b} Morel, T., Miglio, A., Lagarde, N., et
al.\ 2014, \aap, 564, AA119 

\bibitem[Mosser et al.(2010)]{mosser10} Mosser, B., Belkacem, K., Goupil, M.-J.,
et al.\ 2010, \aap, 517, AA22 

\bibitem[Mosser et al.(2011)]{mosser11} Mosser, B., Barban, C., Montalb{\'a}n,
J., et al.\ 2011, \aap, 532, AA86 

\bibitem[Neves et al.(2014)]{neves14} Neves, V., Bonfils, X., Santos, N.~C., et
al.\ 2014, \aap, 568, AA121 

\bibitem[Newberg et al.(2012)]{legue} Newberg, H.~J., Carlin, J.~L., Chen, L.,
et al.\ 2012, Galactic Archaeology: Near-Field Cosmology  and the Formation of
the Milky Way, 458, 405 

\bibitem[Nieva \& Sim{\'o}n-D{\'{\i}}az(2011)]{nieva11} Nieva, M.-F., \&
Sim{\'o}n-D{\'{\i}}az, S.\ 2011, \aap, 532, AA2 

\bibitem[Nieva \& Przybilla(2012)]{nieva12} Nieva, M.-F., \& Przybilla, N.\ 2012,
\aap, 539, AA143 

\bibitem[Ochsenbein et al.(2000)]{vizier} Ochsenbein, F., Bauer, P., \& Marcout,
J.\ 2000, \aaps, 143, 23 

\bibitem[{\"O}nehag et al.(2012)]{onehag12} {\"O}nehag, A., Heiter, U.,
Gustafsson, B., et al.\ 2012, \aap, 542, AA33 

\bibitem[Pasquini et al.(2000)]{flames} Pasquini, L., Avila,  G., Allaert, E., et
al.\ 2000, \procspie, 4008, 129 

\bibitem[Pasquini et al.(2012)]{pasquini11} Pasquini, L., Brucalassi, A., Ruiz,
M.~T., et al.\ 2012, \aap, 545, A139 

\bibitem[Pietrinferni et al.(2004)]{basti1} Pietrinferni, A., Cassisi, S.,
Salaris, M., \& Castelli, F.\ 2004, \apj, 612, 168 

\bibitem[Pietrinferni et al.(2006)]{basti2} Pietrinferni, A., Cassisi, S.,
Salaris, M., \& Castelli, F.\ 2006, \apj, 642, 797 

\bibitem[Przybilla et al.(2011)]{przybilla11} Przybilla, N., Nieva, M.-F., \&
Butler, K.\ 2011, Journal of Physics Conference Series, 328, 012015 

\bibitem[Ram{\'{\i}}rez \& Allende Prieto(2011)]{ramirez11} Ram{\'{\i}}rez, I.,
\& Allende Prieto, C.\ 2011, \apj, 743, 135 

\bibitem[Ram{\'{\i}}rez \& Mel{\'e}ndez(2005)]{ramirez05} Ram{\'{\i}}rez, I., \&
Mel{\'e}ndez, J.\ 2005, \apj, 626, 446 

\bibitem[Randich et al.(2013)]{randich12} Randich, S., Gilmore, 
G., \& Gaia-ESO Consortium 2013, The Messenger, 154, 47 

\bibitem[Repolust et al.(2004)]{repolust04} Repolust, T., Puls, J., \& Herrero,
A.\ 2004, \aap, 415, 349 

\bibitem[Rojas-Ayala et al.(2012)]{rojas12} Rojas-Ayala, B., Covey, K.~R.,
Muirhead, P.~S., \& Lloyd, J.~P.\ 2012, \apj, 748, 93 


\bibitem[Sacco et al.(2014)]{sacco14} Sacco, G.~G., Morbidelli, L., Franciosini,
E., et al.\ 2014, \aap, 565, 113 

\bibitem[Searle et al.(2008)]{searle08} Searle, S.~C., Prinja, R.~K., Massa, D.,
\& Ryans, R.\ 2008, \aap, 481, 777 

\bibitem[Siebert et al.(2011)]{rave3} Siebert, A., Williams, 
M.~E.~K., Siviero, A., et al.\ 2011, \aj, 141, 187 

\bibitem[Simmerer et al.(2013)]{simmerer13} Simmerer, J., Feltzing, S., \&
Primas, F.\ 2013, \aap, 556, A58 

\bibitem[Sim{\'o}n-D{\'{\i}}az et  al.(2006)]{simondiaz06} Sim{\'o}n-D{\'{\i}}az,
S., Herrero, A., Esteban, C., \& Najarro, F.\ 2006, \aap, 448, 351 

\bibitem[Sim{\'o}n-D{\'{\i}}az(2010)]{simondiaz10} Sim{\'o}n-D{\'{\i}}az, S.\
2010, \aap, 510, AA22 

\bibitem[Sim{\'o}n-D{\'{\i}}az et al.(2011a)]{simon11a} Sim{\'o}n-D{\'{\i}}az,
S., Castro, N., Herrero, A., et al.\ 2011, Journal of Physics Conference Series,
328, 012021 

\bibitem[Sim{\'o}n-D{\'{\i}}az et al.(2011b)]{simon11b} Sim{\'o}n-D{\'{\i}}az,
S., Castro, N., Garcia, M., Herrero, A., \& Markova, N.\ 2011, Bulletin de la
Societe Royale des Sciences de Liege, 80, 514 

\bibitem[Sim{\'o}n-D{\'{\i}}az et al.(2011c)]{simon11c} Sim{\'o}n-D{\'{\i}}az,
S., Garcia, M., Herrero, A., Ma{\'{\i}}z Apell{\'a}niz, J., \& Negueruela, I.\
2011, Stellar Clusters \& Associations: A RIA Workshop on Gaia, 255 

\bibitem[Sim{\'o}n-D{\'{\i}}az et al.(2015)]{simon15} Sim{\'o}n-D{\'{\i}}az, S.,
Negueruela, I., Ma{\'{\i}}z Apell{\'a}niz, J., et al.\ 2015, Highlights of
Spanish Astrophysics VIII, 576 

\bibitem[Smiljanic et al.(2014)]{smiljanic14} Smiljanic, R., Korn, A.~J.,
Bergemann, M., et al.\ 2014, \aap, 570, AA122 

\bibitem[Smith \& Dworetsky(1993)]{smith93} Smith, K.~C., \& Dworetsky, 
M.~M.\ 1993, \aap, 274, 335 

\bibitem[Soubiran et al.(2013)]{soubiran13} Soubiran, C., Jasniewicz, G., Chemin,
L., et al.\ 2013, \aap, 552, A64 

\bibitem[Sousa et al.(2014)]{sousa14} Sousa, S.~G., Santos, N.~C., Adibekyan, V.,
et al.\ 2014, \aap, 561, AA21 

\bibitem[Steinmetz et al.(2006)]{rave1} Steinmetz, M.,  Zwitter, T., Siebert, A.,
et al.\ 2006, \aj, 132, 1645 

\bibitem[Stetson(1987)]{daophot1} Stetson, P.~B.\ 1987, \pasp,  99, 191 

\bibitem[Stetson(1992)]{daophot2} Stetson, P.~B.\ 1992,  Astronomical Data
Analysis Software and Systems I, 25, 297 

\bibitem[Taylor(2005)]{topcat} Taylor, M.~B.\ 2005, Astronomical Data Analysis
Software and Systems XIV, 347, 29 

\bibitem[Tang et al.(2014)]{parsec3} Tang, J., Bressan, A., Rosenfield, P., et
al.\ 2014, \mnras, 445, 4287 

\bibitem[Thompson et al.(2008)]{thompson08} Thompson, H.~M.~A., Keenan, F.~P.,
Dufton, P.~L., et al.\ 2008, \mnras, 383, 729 

\bibitem[Thygesen et al.(2012)]{thygesen12} Thygesen, A.~O., Frandsen, S.,
Bruntt, H., et al.\ 2012, \aap, 543, AA160 


\bibitem[Valentini et al.(2013)]{valentini13} Valentini, M., Morel,  T., Miglio,
A., Fossati, L.,  \& Munari, U.\ 2013, European Physical Journal Web of
Conferences, 43, 03006 

\bibitem[VandenBerg et al.(2006)]{victoria} VandenBerg, D.~A., Bergbusch, P.~A.,
\& Dowler, P.~D.\ 2006, \apjs, 162, 375 

\bibitem[Villanova et al.(2010)]{villanova10} Villanova, S., Randich, S.,
Geisler, D., Carraro, G., \& Costa, E.\ 2010, \aap, 509, AA102 

\bibitem[von Braun et al.(2011)]{vonbraun11} von Braun, K., Boyajian, T.~S.,
Kane, S.~R., et al.\ 2011, \apjl, 729, LL26 

\bibitem[von Braun et al.(2012)]{vonbraun12} von Braun, K., Boyajian, T.~S.,
Kane, S.~R., et al.\ 2012, \apj, 753, 171 


\bibitem[Walborn \& Fitzpatrick(1990)]{walborn90} Walborn, N.~R., \&
Fitzpatrick, E.~L.\ 1990, \pasp, 102, 379 

\bibitem[Worley et al.(2012)]{worley12} Worley, C.~C., de Laverny, P.,
Recio-Blanco, A., et al.\ 2012, \aap, 542, AA48 

\bibitem[Yadav et al.(2008)]{yadav08} Yadav, R.~K.~S., Bedin, L.~R., Piotto, G., 
et al.\ 2008, \aap, 484, 609 


\bibitem[Zwitter et al.(2008)]{rave2} Zwitter, T., Siebert,  A., Munari, U., et
al.\ 2008, \aj, 136, 421 




\end{thebibliography}
\end{document}